\documentclass[aps,pra,twocolumn]{revtex4-2}
\usepackage{epsfig,newlfont,amssymb,amsfonts,amsmath,bm,subfigure,palatino,mathtools,amsthm,braket,soul,enumitem,color,graphics,graphicx,times,physics,bbold,comment}

\def\be{\begin{equation}}
\def\ee{\end{equation}}
\def\ba{\begin{eqnarray}}
\def\ea{\end{eqnarray}}
\def\la{\langle}
\def\ra{\rangle}

\usepackage{amsmath}
\usepackage{amssymb}
\usepackage{color, xcolor}
\usepackage{amsthm}
\usepackage{hyperref}
\usepackage{cleveref}
\usepackage{diagbox}
\hypersetup{colorlinks, citecolor=blue, filecolor=blue, linkcolor=blue, urlcolor=blue}
\usepackage[normalem]{ulem}
\usepackage{float}
\usepackage{multirow}
\newtheorem{theorem}{Theorem}
\newcolumntype{P}[1]{>{\centering\arraybackslash}p{#1}}
\graphicspath{{Figures/}{Dynamics/}}

\begin{document}
\title{
Predicting topological quantum phase transition from dynamics via multisite entanglement }
\author{Leela Ganesh Chandra Lakkaraju$^1$, Sudip Kumar Haldar$^{1,2,3}$,  Aditi Sen(De)$^1$}
\affiliation{$^1$Harish-Chandra Research Institute, A CI of HBNI, Chhatnag Road, Jhunsi, Prayagraj - 211019, India\\
$^2$Physics Department, SRM University Delhi-NCR, Rajiv Gandhi Education City, Sonipat 131029, India.\\
$^3$Department of Physics and Material Science $\&$ Engineering, Jaypee Institute of Information Technology,
Noida 201304,}

\begin{abstract}
    An exactly solvable Kitaev model in a two-dimensional square lattice exhibits a topological quantum phase transition which is different from the symmetry-breaking transition at zero temperature.  When the ground state of a nonlinearly perturbed Kitaev model with different strengths of perturbation taken as the initial state is quenched to a pure Kitaev model, we demonstrate that various features of the dynamical state, such as the Loschmidt echo and time-averaged multipartite entanglement, can determine whether the initial state belongs to the topological phase or not. Moreover, the derivatives of the dynamical quantifiers can faithfully identify the topological quantum phase transition, which is present at equilibrium. When the individual qubits of the lattice interact with the local thermal bath repeatedly, we observe that block entanglement in dissipative dynamics can nevertheless distinguish the equilibrium phases from which the system starts evolution.  
\end{abstract}
    
\maketitle

\section{Introduction}
\label{sec:intro}

Phase transitions, which occur when a system parameter crosses a critical value in a condensed matter system, are characterized by a sharp change in behavior.  While the conventional phase transitions are due to the onset of thermal fluctuation after a critical temperature,  there can be a quantum phase transition (QPT) solely driven by the quantum fluctuations that occur by tuning the system parameter \cite{sachdev_2011}. Moreover, it was shown in recent years that in the quench dynamics, quantum critical points can be linked to the non-analytic behavior of physical quantities with time which is also referred to as the dynamical quantum phase transition (DQPT) \cite{Krish2004, Aditi2005, Heylrev, heyl_original, arnab_dqpt_critical1, arnab_dqpt_critical3, jafari_dqpt2}. Both in static and dynamical scenarios, it has been pointed out that multipartite entanglement measures \cite{HoroRMP} can be used as a marker of QPTs and DQPTs present in quantum spin models \cite{Wei05, Anindya14, ultracold-review, fazio-rev, stavPRB}. Due to advancements on the experimental front,  such spin models can nowadays be realized and manipulated in a controlled way using trapped ions \cite{HAFFNER2008155, Duan10}, cold atoms in optical lattices \cite{ultracold-review}, and superconducting qubits \cite{Kaur21}. 

On the other hand, it was shown that systems with topologically ordered states possess several unique characteristics like robustness under local perturbation which is, in general, absent in other phases of a many-body system \cite{Wen95, Wen02}. Moreover, such states cannot be characterized by any local order parameter, and hence, QPTs from topologically ordered states cannot be understood by conventional theories based on the divergence in local order parameters. Therefore, a novel approach is required to investigate the underlying characteristics of the ground state in these systems,  especially, the QPT from a topological phase to another phase, known as the topological quantum phase transition (TQPT) which has also been extensively studied using both analytical and numerical techniques \cite{Hamma2008, Trebst07, PRB77}.    
 In this context, the Kitaev toric code is an example of a topologically ordered state that undergoes a second-order quantum phase transition \cite{Kitaev02, KITAEV20032, KITAEV20062}. 
The ground states of the modified Kitaev models also change phase from the topologically ordered phase to a nontopological one, thereby exhibiting a TQPT \cite{Trebst07, Claduio2008, Wu12}. 

Further, topologically ordered states are of particular interest in quantum information processing tasks \cite{Kitaev02, RevModPhys.80.1083} which include quantum communication, quantum computation, and quantum error correcting codes since they are resilient to local perturbation \cite{Levin2006, Kitaev2006, Jiang2012, Yao2010, Abasto2008, PRB77, Chen2010, Zhang2022, Schotte2019}. More specifically, the toric code can be used as quantum memory to store quantum information due to its robustness against local and thermal errors \cite{Kitaev02}. Several information theoretic quantities, such as block entanglement entropy \cite{PRB77, entropy_tqpt2, entropy_tqpt3}, multipartite entanglement \cite{Zarei2022}, quantum discord \cite{Chen2010}, and the Fisher information \cite{Zhang2022} of the ground state in the Kitaev code and quantities derived from them have been shown to be useful to detect the TQPT. Note, however, that the bipartite reduced states possess vanishing entanglement, thereby making them incapable of detecting a TQPT.  A very recent work \cite{amit22} demonstrated that the TQPT may be distinguished using localizable entanglement obtained from the dynamical state of the Kitaev code in the presence of a parallel magnetic field which is influenced by Markovian and non-Markovian dephasing noise. Topological quantum criticalities were studied under non-equilibrium conditions \cite{arnab_tdqpt_critical2,jafari_longrange_kitaev_dqpt2, jafari_utkarsh_longrange_kitaev_dqpt1}, and it was shown that under Floquet or periodic driving, dynamical states can also predict  \cite{jafari_floquet_dqpt1,jafari_floquet_dqpt2, jafari_floquet_dqpt3, jafari_floquet_dqpt4} equilibrium critical points  \cite{jafari_critical_dqpt} including topologically ordered critical points.  \cite{floquet_milad1, jafari_floquet_nonhermitian_topological_dqpt}.  



In our work, we examine the nonlinearly perturbed Kitaev code, which was demonstrated to undergo a topological quantum phase transition with the adjustment of the perturbation strength at zero temperature \cite{Claduio2008, Dusuel2011, Zarei2015, Zarei2019, Hamma2008, Vidal2009}.
Here, we address the question of whether the characteristics of the evolved state are capable of indicating the occurrence of a topological quantum phase transition in equilibrium, which we refer to as a topological dynamical quantum phase transition (TDQPT). In this respect, it is important to stress that there are quantum spin models, such as the quantum $XY$ model with uniform and alternating magnetic fields, in which the dynamical states cannot faithfully recognize phase transitions that occurred at equilibrium \cite{stavPRB}. However, using this perturbed Kitaev model, the equilibrium topological transitions can be successfully predicted from the dynamics. Specifically, the initial state is taken to be the nonlinearly perturbed Kitaev model with different strengths of perturbation, and the system is then quenched to an original Kitaev model. We illustrate that the rate function originated from Loschmidt echo (LE), a conventional measure for detecting DQPT , shows non-analyticity in time when the quenching is performed across the quantum topological critical point. However, any such non-analyticity is not observed when both the initial and evolved states belong to the same phase.


We demonstrate that the genuine multipartite entanglement measure, quantified by the generalized geometric measure (GGM) \cite{ggm_aditi} and block entanglement of the dynamical state can successfully determine the TQPT present in the ground state, even though entanglement has not yet been established as a quantifier for identifying DQPT (cf. \cite{stavPRB}). In particular, both the time-averaged GGM and block logarithmic negativity \cite{neg4, neg5} of the evolved state change from concave to convex with the variation of the strength in the nonlinear perturbation at the phase transition point, resulting in non-analytic behavior in their derivatives. Further, we observe that if both the initial and final Hamiltonians belong to the topologically ordered phases, the evolved state possesses a substantial amount of average multipartite entanglement. Going beyond the unitary dynamics, our studies also reveal that when the entire system is affected by the local environment, the time-averaged block entanglement decays, although the behavior of entanglement can still predict the topological critical point.


The organization of this paper is as follows. In Sec. \ref{sec:models}, we introduce the nonlinearly perturbed Kitaev code as well as topological criticalities in static scenarios and also describe the evolution due to sudden quench. The physical quantities that we apply to detect the quantum phase transition in dynamics are discussed in Sec. \ref{sec:measure}. In Sec. \ref{sec:dqptecho}, we present the patterns of the Loschmidt echo and the multipartite entanglement of the evolved states and illustrate that their dynamical behavior can detect the TQPT at equilibrium. When the local noise affects all the sites in the lattice, the entanglement of the dynamical state is still capable of identifying quantum criticality as shown in Sec. \ref{sec:criticaldecoh}. We summarize in Sec. \ref{sec:conclu}.

\section{Toric Code with nonlinear perturbation}
\label{sec:models}

Let us first introduce the Hamiltonian that we will use to demonstrate the topological dynamical quantum phase transition. Specifically, we identify the parameters which may be used to observe the dynamical quantum phase transition.  

\subsection{Nonlinearly perturbed Kitaev code}

\begin{figure}
    \centering
    \includegraphics[width=\linewidth]{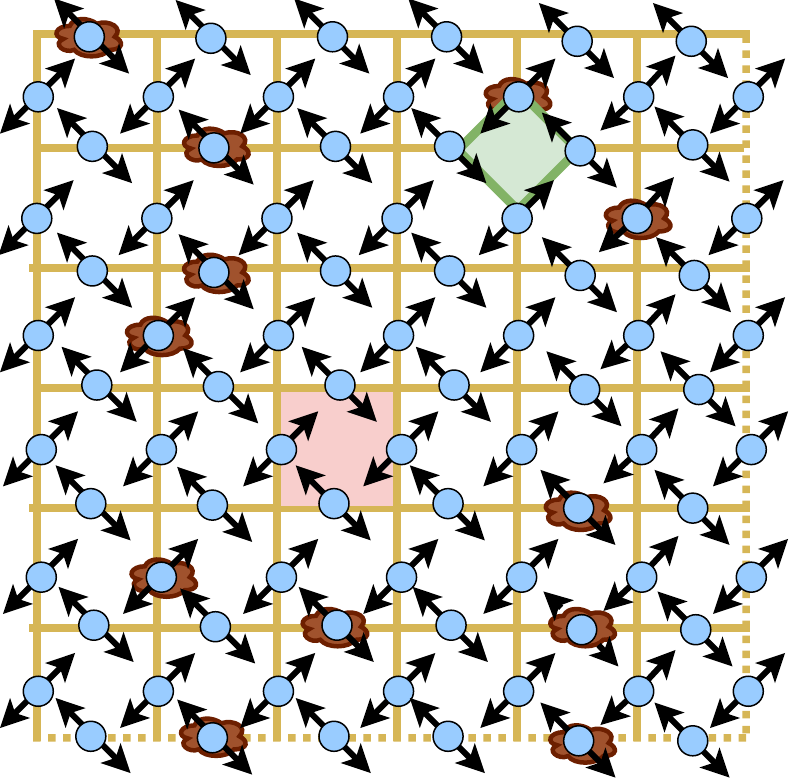}
    \caption{Schematic diagram representing the Kitaev toric code with the periodic boundary condition (which is shown by dotted lines at the boundary). The star and plaquette operators, \(A_v\) and \(B_p\), are marked in green and pink, respectively. Moreover, the clouds represent the local noise that acts on each qubit or some of the qubits of the entire system.}
    \label{fig:schematic}
\end{figure}

For the present work, we consider a deformed Kitaev toric code with a nonlinear perturbation on a two-dimensional (2D) square lattice consisting of vertices and plaquettes with spin-\(1/2\) particles located on each edge of the square grid of the lattice. The Hamiltonian in this case reads \cite{Claduio2008, Zarei2022}
\begin{equation}\label{H}
\hat{H}_{NLTC}(\beta)=-\sum_v \hat{A}_v -\sum_p \hat{B}_p+\sum_v e^{-\beta \sum_{i \in v} \hat{\sigma}_i^z}, 
\end{equation}
where \(\beta >0\), with \(\beta =0\) representing the original Kitaev model, and \(\hat{\sigma}^k\) (\(k=x, y, z\)) is the Pauli matrix.  It was shown that the above model exhibits a second-order topological quantum phase transition \cite{Claduio2008} as the system parameter \(\beta\) is tuned across the critical value,  \(\beta_{critical} = \frac{1}{2}\ln (\sqrt{2} +1) = 0.4407\) (see Appendix \ref{appendix:deri} for details). Here, $\hat{A}_v$ and $\hat{B}_p$ represent the star and plaquette  operators, respectively  which are defined as the tensor products of Pauli operators, $\hat{\sigma}_i^x$ and $\hat{\sigma}_i^z$, acting on an individual spin-$\frac{1}{2}$ particle, 
\(\hat{A}_v = \prod_{i \in v} \hat{\sigma}_i^x\) and \(\hat{B}_p = \prod_{i \in p} \hat{\sigma}_i^z\) (see Fig. \ref{fig:schematic}),  
with \(N/2\) being the total number of vertices.

As the name suggests, the star operator acts on four qubits situated around a vertex of the lattice while the action of the plaquette operator is again on four qubits on the bond around a plaquette of the $2$D lattice. Similarly, the perturbative exponential term in Eq.~(\ref{H})  simultaneously acts on the four qubits belonging to a particular star. Therefore, the toric code Hamiltonian[Eq. 
 (\ref{H})] is composed of four-body nonlocal operators.

From the definition, all stars and plaquette operators commute with each other. Furthermore, $\hat{A}_{\nu}$ are generators of an Abelian group whose elements can be represented by a loop configuration. All possible trivial loops can be generated by the combination of $\hat{A}_{\nu}$ on each star. An Abelian group \cite{Zanardi2007} comprising of all possible loops helps in identifying the underlying structure of the ground state; i.e., the element of the group is $a_{\left\{r_1, r_2, \ldots, r_{N / 2}\right\}}=\hat{A}_1^{r_1} \cdot \hat{A}_2^{r_2} \ldots \hat{A}_{\frac{N}{2}-1}^{r_{\frac{N}{2}-1}}$, where $r_i$ takes a value of $1$ or $0$ depending on the corresponding star operator being active on a site $i$ or not. 
The underlying lattice becomes a torus as the periodic boundary conditions are imposed on both the horizontal and vertical edges of the lattice. The torus structure increases the number of ground states to four linearly independent states which satisfy the toric code. In particular, the centers of two non-trivial loops match the centers of the tube and the torus, respectively.

Since Eq. (\ref{H}) reduces to the analytically solvable Kitaev toric code with some energy shifts for $\beta=0$, the ground state (GS) properties are already known \cite{KITAEV20062} in this limit. One of the ground states in the 
four-dimensional GS manifold can be expressed as~ \cite{Claduio2008}
\begin{equation} 
    |GS\rangle = \prod_v (1+\hat{A}_v) |0\rangle^{\bigotimes N}.
    \label{Eq.Kitaev-Ground}
\end{equation}
Here, $N$ is the total number of spins and $|0\rangle^{\bigotimes N}$ represents a fully magnetized state with all spins pointing up. Therefore, one can immediately identify Eq.~(\ref{Eq.Kitaev-Ground}) as the ground state of Eq.~(\ref{H}) in the limit $\beta \rightarrow 0$.    

In the other extreme limit, i.e., $\beta \rightarrow \infty$, the ground state of Eq.~(\ref{H}) becomes fully magnetized, which suggests that the system exhibits a topological quantum phase transition from a topological phase to a magnetized phase as $\beta$ is varied from $\beta=0$ to $\beta \rightarrow \infty$.

The exact ground state of the system can be analytically obtained ~\cite{Claduio2008} as
\begin{equation} \label{gndstate}
    |GS(\beta)\rangle = \frac{1}{\sqrt{Z(\beta)}}\sum_{a \in G} \exp^{\beta\sum_i \hat{\sigma}_i^z(a)} a|0\rangle,
\end{equation}
where $a \in G$ refers to the loop operators from the Abelian group $G$, $Z(\beta)= \sum_{a \in G} \exp^{\frac{\beta}{2}\sum_i \hat{\sigma}_i^z(a)}$, and $\hat{\sigma}^z_i(a) =\mp 1$ depending on whether the spin $i$ has an intersection with the loop operator $a$ or not.
The static properties of \(\hat{H}_{NLTC} (\beta)\) can help us to fix the initial state and the quenching Hamiltonian.  

Before we start the investigation, let us make a note of the recent efforts to study the features like entanglement of these kinds of Hamiltonian in laboratories. 
Instead of realizing the Hamiltonian and measuring its properties, it is more common to prepare the topologically ordered ground state of the toric code using quantum circuits with different numbers of qubits. Specifically, a quantum circuit comprising $31$ superconducting qubits on a Sycamore quantum processor is used to prepare the toric code ground state ~\cite{Satzinger}. More precisely,  it is implemented by simultaneous application of suitable combinations of the Hadamard gate on one of the four qubits in a plaquette following a CNOT gate on the other qubits.  Other implementations of these types of ground state include trapped Rydberg atoms ~\cite{Vishwanath_PRX, Vishwanath_science, Auger} and polarized photons~\cite{Pachos_2009}.  In the small $\beta$ limit, such set-ups with a magnetic field having strength $\beta$ in the $z$ direction have the potential to investigate the properties of the ground and dynamical states of the Hamiltonian, \(H_{NLTC}\).

\subsection{Quench across the critical point}\label{quench}

A sudden change in parameters under evolution, more popularly known as quantum quench or sudden quench turns out to be an important tool for studying the non-equilibrium properties of the system under consideration. It has been established that the ground state of the toric code is resilient to local perturbations. We consider a study in which the system is no longer at equilibrium and actively undergoes evolution \cite{Fazio2009}.

To achieve the goal of mimicking equilibrium physics, especially the TQPT from the dynamical state, the initial state is chosen to be the ground state of  $H_{NLTC}(\beta_0)$; i.e., at \(t=0\), \(|GS\rangle\) is taken as the initial state for dynamics. After the sudden quench in which \(\beta_0\) is abruptly changed to \(\beta_1\), the system evolves according to  the  Hamiltonian $H_{NLTC}(\beta_1)$.

To identify TQPT through evolution, we ensure that \(\beta_0\) and $\beta_1$ are taken from either the same phase or different phases \cite{Heylrev, heyl_original}. 
The evolved state takes the form 
\begin{equation}
    |\psi(\beta_0, \beta_1, t)\rangle = U(\beta_1, t)|\psi_0(\beta_0)\rangle = e^{-iH_{NLTC}(\beta_1)t}|\psi_0(\beta_0)\rangle,
\end{equation}
where $|\psi_0(\beta_0)\rangle$ is the ground state of the Hamiltonian $H_{NLTC}(\beta_0)$. 
For our investigation, the initial state is chosen with \(\beta_0\neq 0\) while the post-quenched Hamiltonian is always considered to be the original Kitaev code, i.e., $\beta_1 = 0$. 
In particular, the  time evolution (unitary) operator $U(t)$ is given by
\begin{align}
    U(\beta_1 = 0, t) &= U(t)
    = e^{-i H_{toric} t} \nonumber \\ &\equiv e^{i\sum_v \hat{A}_v t } = \Pi_{l \in v} \cos(t) \mathcal{I}_{16} + i \sin(t) \hat{A}_l ,
    \label{eq.unitary}
\end{align}
where $\mathcal{I}_{16}$ refers to the identity matrix in the $16$-dimensional complex Hilbert space. Note that the time evolution operator corresponding to the plaquette operator $(\hat{B}_p)$ does not play a role because the $\hat{\sigma}^z$-operator acting on the initial state only contributes to the phase of the state and hence only the star operator $(\hat{A}_v)$ contributes in Eq.~ (\ref{eq.unitary}). It is important to note that the above unitary operator possesses nonlocal interaction \cite{Fazio2009, Zhang2022}, thereby having the capability to produce entanglement. Specifically, the toric code is a four-body operator as mentioned above, with coefficients depending on time. 

\section{Measures used for detecting Criticalities in Evolution}
\label{sec:measure}

Let us briefly discuss the quantifiers that we use to detect the TQPT in dynamics, which we call the topological dynamical quantum phase transition (TDQPT). 

{\it Loschmidt echo and rate function.} First, we employ the conventional DQPT detector, the Loschmidt echo  \cite{heyl_original, Heylrev}  which is computed to distinguish the equilibrium phases from the dynamical state. 
It is defined as
 $\mathcal{L}(t) =| \langle \psi_t | \psi_0\rangle|^2$, 
where \(| \psi_0\rangle\) and \(| \psi_t\rangle = U(t) | \psi_0\rangle\) are the initial and evolved states respectively. The evolved state after the action of $U(t)$ reads as 
\begin{align}
 |\psi_t \ra & \equiv |\psi(\beta_1 = 0, \beta_0, t) \ra  =  U(t)|GS(\beta_0) \ra \nonumber \\
    &= \prod_{l \in v} \cos(t) \mathcal{I}_{16} + i \sin(t) \hat{A}_l  \nonumber \\
    & \times \Big ( \frac{1}{\sqrt{Z(\beta)}}\sum_{a \in G} \exp^{\beta\sum_i \hat{\sigma}_i^z(a)} a|0\rangle \Big ) \nonumber \\
    &= \frac{1}{\sqrt{Z(\beta)}}  \sum_{\substack{ \{a_1, a_2\} \in G \\ r = \{0.. \frac{N}{2}\}}} \cos(t)^{\frac{N}{2}-r} (i \sin(t))^r \nonumber \\
    & \times \exp^{\beta\sum_i \hat{\sigma}_i^z(a_1 a_2)} a_1 a_2|0\rangle, \label{eq:toric_quench_evolution}
\end{align} where $a_1 a_2$ also belongs to $G$ due to the closure property of the Abelian group.  This ensures that the evolved state retains a structure similar to that of the ground state, making the calculation of $\mathcal{L}(t)$ and the corresponding rate function, defined below, possible. 


It has been observed that in the case of a quantum transverse Ising chain, the evolved state becomes completely orthogonal to the initial state if the quench is performed to a phase other than the initial one \cite{heyl_original}. To detect the existence of such zeros, the logarithm of $\mathcal{L}(t)$, known as the Loschmidt rate, given by
$\Lambda(t) = \lim_{N \to \infty} \frac{-1}{N}{\ln[\mathcal{L} (t)]} =  \lim_{N \to \infty} \frac{-2}{N}{\ln |\langle \psi_t | \psi_0\rangle}|$, is used.
The non-analytic behavior of the rate function with time is argued to be analogous to the behavior of the free energy in the classical phase transition \cite{Heylrev} if we replace $i \times t$ in the evolution operator with the inverse temperature in the partition function. 

\begin{figure*}
\centering
\includegraphics[width=\linewidth]{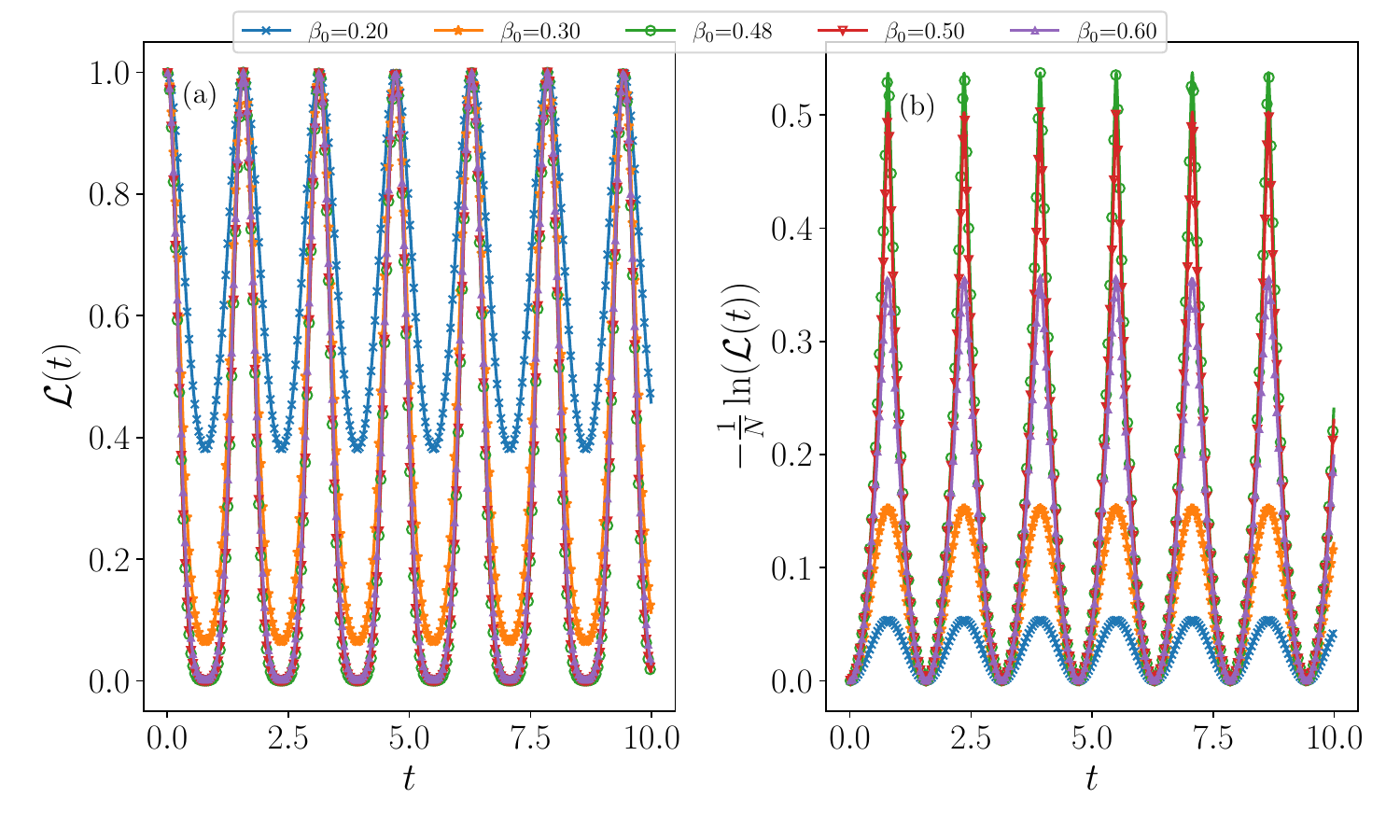}
\caption{(a) Loschmidt echo  (ordinate) of the nonlinearly perturbed Kitaev model vs time, $t$ (abscissa) for different choices of parameters for the ground state as initial states, \(\beta_0\). The sudden quench is performed with the Hamiltonian having \(\beta_1 =0\). The initial and final Hamiltonians are in different phases when \(\beta_0 > \beta_{critical} = 0.4407\), while they are in the same phase for \(\beta_0 < \beta_{critical} = 0.4407\).   (b) Rate function \(\Lambda(t)\) (vertical axis)  with respect to time (horizontal axis). The system size is chosen to be \(18\).  All the axes are dimensionless. }
\label{fig:loschmidt_echo}
\end{figure*}

Specifically, it was shown that uniformly spaced kinks appear in the time evolution of the rate function for the transverse Ising spin model when the initial 
Hamiltonian and the quenched Hamiltonian belong to different phases while such kinks are absent if they are chosen from the same phase. Notice, however, that many exceptions to this detection process are also reported \cite{Vajna14, Sirker14,gurarie_pra_2019, stavPRB, nandi_prl_2022, Jafari_dqpt1}. However, as we will illustrate in the next section, both \(\mathcal{L}(t)\) and \(\Lambda(t)\) are capable of identifying the TQPT from the dynamics.



{\it Entanglement measures.} Let us define two entanglement measures, namely the genuine multipartite entanglement measure, which is computed by exploiting the geometric structure of states \cite{ggm_aditi, ggm1, Wei05, ggm3, ggm4, ggm_mixed_otfried,aditi_ggm_mixed}, and block entanglement, which is computed by dividing the entire system into two equal blocks. 
 


{\it Genuine multipartite entanglement.} A pure state is genuinely multipartite entangled if it is not separable in any bipartition (after dividing the entire system into two blocks). Moreover, we know  that  
all separable states form a closed and convex set. It gives rise to the possibility of 
measuring the entanglement of a given state by calculating the distance between the set of separable states and the given state.  
By denoting the set of non-genuinely multipartite entangled states as $\Lambda_G$, the GGM 
is defined as
 \( \mathcal{G} (|\psi\rangle)=1-\max_{|\phi\rangle \in \Lambda_G} |\langle\phi \mid \psi\rangle|^2,\, \, \) such that it measures the distance of a given state from the closest biseparable state \(|\phi\ra\). 

By using Schmidt decomposition of a pure state, the GGM reduces to a simple form as \cite{ggm_aditi}
\begin{align}
&& \mathcal{G}(|\psi\rangle) =1-\max \left\{\lambda_{i_1 \text { :rest }}^{\max }, \lambda_{i_1 i_2 \text { :rest }}^{\max _1}, \ldots, \lambda_{i_1 i_2 \ldots i_M: \mathrm{rest}}^{\max } \mid\right.\nonumber \\
&&i_1, i_2, \ldots, i_M \left.\in\{1,2, \ldots \frac{N}{2}\} ; i_k \neq i_l ; k, l \in\{1,2, \ldots M\}\right\},\nonumber\\
\end{align}
where $\lambda^{\max}_{i_1 i_2 \ldots i_M}$ is the largest eigenvalue of the reduced density matrix, \(\rho_{i_1 i_2 \ldots i_M}\) corresponding to a bipartition \(M:rest\). Also, in our case, $N$ is always even since an odd number of spins cannot sit on a torus. 
Although it may seem that one has to calculate all possible bipartitions, which requires  computing the maximum eigenvalues of  $\sum_{i = 1}^\frac{N}{2} \dbinom{N}{i}=\frac{1}{2}\times(2^{n}+\dbinom{N}{\frac{N}{2}}-2)$ matrices, we will prove that this is not the case in the next section.



\begin{figure*}
    \includegraphics[width=\linewidth]{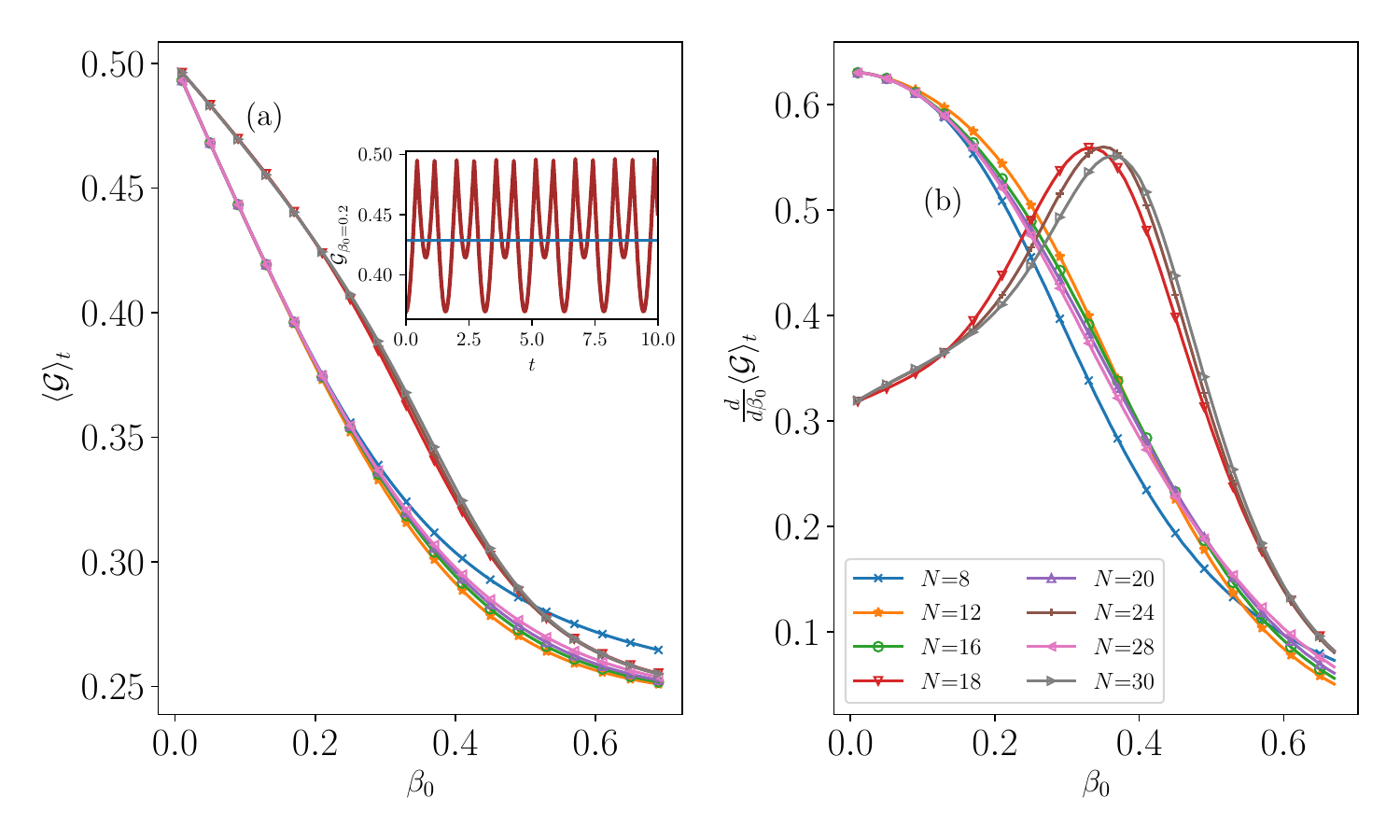}
    \caption{(a) Time-averaged GGM, $\langle \mathcal{G} \rangle_t$ (ordinate) of the evolved state of \(\hat{H}_{NLTC}\) against $\beta_0$ (abscissa)  for different sizes of $N$. (Inset) Behavior of GGM with $t$ where $\beta_0 = 0.2$ and $\beta_1 = 0$ and the (blue) line represents the time-averaged GGM (\(\langle \mathcal{G}\rangle_t\)). (b) The first derivative of $\langle \mathcal{G} \rangle_t$ (y-axis) with respect to $\beta_0$ (x-axis).  A peak in $\frac{d\langle \mathcal{G} \rangle_t}{d \beta_0}$  is observed at $\beta_0$ close to the TQPT. This indicates that the $\langle \mathcal{G} \rangle_t$ changes from concave to convex at the transition point. The value of $\beta_0$ reaches the exact value of the TQPT as $N$ increases.  All the axes are dimensionless. 
    }
    \label{fig:ggm_dggm}
\end{figure*}
\textit{Block entanglement.} In addition to the GGM, 
we also study the block entanglement of the time-evolved multipartite state in which entanglement is computed
after dividing the partition of the entire system into two halves, i.e., in the bipartition, $\frac{N}{2}:\frac{N}{2}$. The entanglement is quantified via logarithmic negativity (LN) \cite{neg4}, which is based on the partial transposition criteria \cite{peres, horodecki1996}. For an arbitrary state \(\rho_{AB}\), LN can be defined as \cite{neg4} 
$E_{LN}(\rho_{AB}) = \log_2||\rho_{AB}^{T_A}||$,
where $||\cdot||$ represents the trace-norm and \(\rho_{AB}^{T_A}\) denotes the partial transposition of \(\rho_{AB}\) with respect to the party \(A\). In the case of  pure states, 
 $|\psi \rangle = \sum_\alpha c_\alpha|A_\alpha\rangle|B_\alpha\rangle$,    the negativity, \(\mathcal{N}(\rho) = \frac{||\rho_{AB}^{T_A}||-1}{2}\) reduces to  
\(\mathcal{N}(\rho)=\frac{1}{2}\left[\left(\sum_\alpha c_\alpha\right)^2-1\right]\) using Schmidt decomposition  \cite{neg4}.
By considering the dynamical state, \(|\psi_t\rangle\) with \(N\) parties, we calculate the logarithmic negativity  
in the bipartition \(N/2: N/2\), which we denote \(E_{LN}(|\psi_t\ra_{\frac{N}{2}: \frac{N}{2}})\). 

\section{Detection of Topological Criticalities via Entanglement}
\label{sec:dqptecho}

To uncover the topological quantum phase transition from the dynamics of the system, we adopt three quantities as defined in the preceding section. We start with the commonly used quantifier for the DQPT, the Loschmidt echo, and the rate function. 


\begin{figure*}
    \centering
    \includegraphics[width=\linewidth]{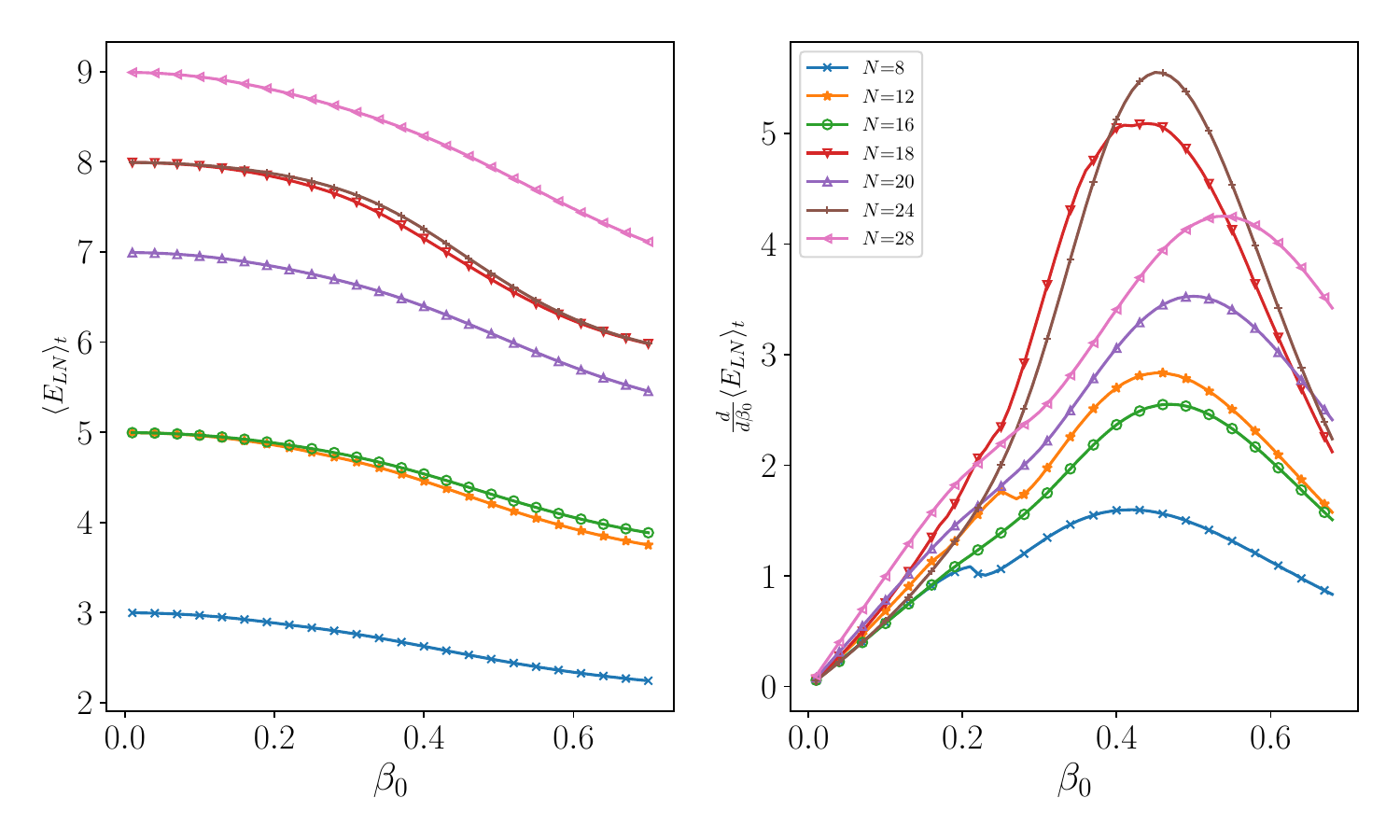}
    \caption{(a) Behavior of the time-averaged block logarithmic negativity $\langle E_{LN} \rangle_t$ (ordinate) with respect to $\beta_0$ (abscissa) in the modified Kitaev model for various system-sizes $N$.  LN is computed by dividing the \(N\)-party state into two equal blocks.  (b)  $\frac{d\langle E_{LN} \rangle_t}{d\beta_0}$ ($y$-axis) vs $\beta_0$ ($x$-axis). The peak at \(\beta_0\) signals the topological quantum phase transition occurred at zero temperature.   All the axes are dimensionless.}
    \label{fig:log_neg_closed}
\end{figure*}


{\it Behavior of the Loschmidt echo and kink in the rate function.} 
The initial state is the ground state of the Hamiltonian, \(\hat{H}_{NLTC}(\beta_0)\) with the values of $\beta_0$ chosen from both  \(\beta_0<\beta_{critical}\) and also \(\beta_0> \beta_{critical}\). As mentioned before, the unitary operator involves the Kitaev Hamiltonian, i.e., \(\hat{H}_{NLTC} (\beta_1 =0)\). Hence when \(\beta_0> \beta_{critical}\), the initial state is the ground state of the system with a paramagnetic phase, thereby belonging to a different phase than the quenching Hamiltonian, while for \(\beta_0<\beta_{critical}\), both the post- and pre-quenched Hamiltonian belong to the same topologically ordered phase.


We observe that when $\beta_0 > \beta_{critical}$, the corresponding Loschmidt echo reaches zero for certain values of \(t\) which are equally spaced in \(t\) like in the transverse Ising model, as depicted in Fig. \ref{fig:loschmidt_echo}(a). On the other hand, for $\beta_0 < \beta_{critical}$, since the quenched Hamiltonian and the ground state lie in the same phase,  \(\mathcal{L}(t)\) never vanishes.
Therefore, the Loschmidt echo, involving both the initial and dynamical states, clearly recognizes the topological quantum phase transition occurred at zero temperature. As argued before, such behavior becomes more evident when one considers the rate function \(\Lambda(t)\). Specifically, \(\Lambda(t)\) demonstrates a kink exactly at those times when \(\mathcal{L}(t)\) vanishes with $\beta_0 < \beta_{critical}$  [see Fig. \ref{fig:loschmidt_echo} (b)]. This signifies that both quantities are capable of predicting the TQPT from the dynamical state. 


\subsection{Detection of
the topological critical point using time-averaged multipartite entanglement}
\label{sec:entcritical}

Beyond the typical indicator of the DQPT, let us demonstrate that entanglement measures, especially multipartite entanglement measures of the evolved state carry the signature of the TQPT (see \cite{stavPRB} for different spin models). It has already been established that both bipartite entanglement and multipartite entanglement of the ground state are capable of detecting the quantum phase transition in the spin Hamiltonian, including the topological phase transition considered in this paper (for global entanglement, see Ref. \cite{Zarei2022}). 


To study the behavior of genuine multipartite entanglement of the evolved state (see Fig. \ref{fig:ggm_dggm}), we compute the time-averaged  GGM, denoted by $\langle \mathcal{G} \rangle_t$, when the state evolves under the quench $\hat{H}_{NLTC}(\beta_1 = 0)$ from the ground state of $\hat{H}_{NLTC}(\beta_0 \ne 0)$ as the initial state. 
Before presenting the trends of \(\langle \mathcal{G} \rangle_t\), we first prove the following theorem, which makes the computation of the GGM  simple for this model. 



\begin{theorem}
Eigenvalues from a single-party density matrix of the evolved state only contribute to the maximum involved in GGM. 
\end{theorem}

\begin{proof}

To find the eigenvalues in bipartitions, we are required to trace out some of the parties, say, \(M\). First, notice that the ground state is a superposition of all possible closed loops on the torus.
Let us consider the density matrix corresponding to the ground state, given by
\begin{equation}
    \hat{\rho}(\beta) = \frac{1}{Z(\beta)} \sum_{a,a^\prime \in G} e^{\frac{\beta}{2}(\hat{\sigma}_i^z(a)+\hat{\sigma}_i^z(a^\prime)) }a|0\ldots 0\rangle \langle0 \ldots 0| a^\prime.
\end{equation}
The corresponding reduced density matrix of  $\{i_1, i_2, i_3 \ldots, i_M \}$ parties is obtained by tracing out \(M\) parties as
\begin{eqnarray}
&&\hat{\rho}_{i_1i_2,i_3 \ldots i_M}= \frac{1}{Z(\beta)} \sum_{\left\{\delta_k=0,1 \right\}} \sum_{a, a^{\prime}} e^{\frac{\beta}{2} \sum_i\left[\hat{\sigma}_i^z(a)+\hat{\sigma}_i^z\left(a^{\prime}\right)\right]} \nonumber\\
&& \times\left\langle\delta_1 \delta_2 \ldots \delta_{N-M}|a|0\ldots 0\rangle \langle0 \ldots 0\left|a^{\prime}\right| \delta_1 \delta_2 \ldots \delta_{N-M}\right\rangle.\nonumber\\
\end{eqnarray}
As shown in Refs. \cite{Zarei2022, Claduio2008}, one can prove by contradiction that this state cannot have non-diagonal terms. As described previously, all loops are represented as elements of an Abelian group generated by $A_\nu$. In the ground state, each $a_i|0\rangle^{\otimes N}$ corresponds to the system configuration of $N$ spins in the $i$-loop configuration. For example, no two closed loops can have one  different spin, and hence it is not possible to have 
\begin{equation}
    \begin{array}{l}
\left(\left\langle\delta_1, \delta_2, \ldots, \delta_x, \ldots, \delta_N\right|\right)\left(a_1|0\rangle^{\otimes N}\right) \\
\quad \times\left({ }^{N \otimes}\langle 0| a_2\right)\left(\left|\delta_1, \delta_2, \ldots, \delta_x^{\prime}, \ldots, \delta_N\right\rangle\right) \ne 0,
\end{array}
\end{equation}
when  $x \ne x^\prime$. A similar argument can be made for any reduced density matrices for the ground state which is the initial state during evolution. 

Since the evolution operator involves the Kitaev model, the evolved state can be written in the same basis as the ground state and the corresponding local density matrices are again diagonal by the same logic as the local density matrices for the ground state. The evolved state is written as $|\psi_t \rangle = \sum_{g \in G}  C_i(t) g|0\rangle^{\bigotimes N}$. This can be easily seen by noticing that only the star operators act on the state while the plaquette operators simply generate a global phase. 

Moreover, we note that the \(M\)-party reduced state has eigenvalues \(e_1, \ldots e_{2^M}\), written in decreasing order. It can easily be found that the \((M-1)\)-party reduced state has \(2^{M-1}\) eigenvalues of the form \(e_i + e_{i+1}\) (\(i =1, 2, \ldots, 2^{M}-1\)). This clearly shows that the maximum eigenvalue of \((M-1)\)-party is bigger than that obtained from the \(M\)-party state. Hence, the single-site reduced density matrix has the maximum eigenvalue which contributes to the computation of the GGM for the evolved state. 


\end{proof}

Let us now elaborate on the way we compute the time-averaged GGM. Following the quench as described before, we calculate the GGM at every time step and  then perform averaging over time, indicated as
\begin{equation}
    \langle \mathcal{G} \rangle_t = \frac{\sum_ {t = t_i}^{t_f} \mathcal{G} (|\psi (\beta_0, \beta_1, t)\rangle)}{L}, 
    \label{eq:GGM_avg}
\end{equation}
where $t_{i(f)}$ is the initial (final) time, and $L = \frac{t_f - t_i}{\delta t}$ with $\delta t$ being the step size. For illustration in Fig. \ref{fig:ggm_dggm}, we choose \(t_i\) and \(t_f\) to be \(0\) and \(10\), respectively, while the step size is taken to be \(0.01\). Since $\mathcal{G} (|\psi (\beta_0, \beta_1, t)\rangle)$ includes sine and cosine functions of time, it is an oscillatory function of time [see the inset in Fig. \ref{fig:ggm_dggm}(a)], and therefore, the choice of $t_f$ does not affect the behavior of $\la \mathcal{G} \ra_t$ as long as $t_f$ includes complete oscillations of $\mathcal{G} (|\psi (\beta_0, \beta_1, t)\rangle)$. This implies that the observables corresponding to the entire time-evolved state which is the case for the GGM, in general, display oscillatory behavior while the observables which involve only subsystems described by a mixed state, may oscillate or may saturate to a value corresponding to the thermal state of the subsystem.
The main observations can be summarized as follows. 
\begin{enumerate}

    \item 
For a very low \(\beta << \beta_{critical}\), the time-averaged GGM is very high, almost close to its maximum value. In other words, the initial state should be prepared as the ground state of the Hamiltonian with a very low \(\beta << \beta_{critical}\) to produce a highly genuine multipartite entangled state during dynamics. Thus, for a small $\beta$, both the initial and the post-quenched Hamiltonian are in the topological phase.  Such a behavior can be termed as {\it topological robustness} which persists in the presence of a weak perturbation, \(\beta_0\). 

    \item With the increase of the perturbation, \(\beta_0\), \(\langle \mathcal{G}\rangle_t\) decreases as shown in Fig. \ref{fig:ggm_dggm} (a). When the initial and post-quenched Hamiltonian belong to different phases, i.e., when the initial and quenching Hamiltonian are in the paramagnetic and topological phases, respectively, the time-averaged GGM content is lower than the scenario with both the initial and final Hamiltonians being in the same phase. 

    \item At the topological phase transition point  $\beta=\beta_{critical}$, the curvature of \(\langle \mathcal{G}\rangle_t\) with $\beta$ changes from concave to convex. 
    Therefore, the derivative of the time-averaged GGM with respect to \(\beta_0\), i.e., \(\frac{d\langle \mathcal{G}\rangle_t}{d\beta_0}\), is maximum at \(\beta = \beta_{critical}\) (see Fig. \ref{fig:ggm_dggm} (b)).  However, for certain small system sizes, this is not the case, and the double derivative of \(\langle G\rangle_t\) is maximum at $\beta_{critical}$. 

\end{enumerate}
The above observations strongly indicate that genuine multipartite entanglement of the dynamical state can efficiently signal the topological phase transition at equilibrium.  Moreover, high-multipartite-entanglement content shows the beneficial role of the topologically ordered phase in the deformed Kitaev model and its importance in quantum information processing tasks. 

{\it Block entanglement.} Let us now examine whether other multipartite entanglement measures are also able to reveal the TQPT from the dynamics. Towards that aim, we compute the time-averaged value of LN in the \(N/2:N/2\) bipartition; i.e., we replace \(\mathcal{G}\) by \(E_{LN}\) in the definition of \(\langle \mathcal{G}\rangle_t\) in Eq. (\ref{eq:GGM_avg}). For all system-sizes, the curvature of \(\langle E_{LN} \rangle_t \equiv \langle E_{LN} (|\psi_t\rangle_{N/2: N/2}) \rangle_t \)  
also changes from convex to concave with the variation of $\beta_0$.  The point of inflection indicates the topological phase transition at zero temperature which is prominent in 
the behavior of \(\frac{d\langle E_{LN} \rangle_t}{d\beta_0}\), as shown in Fig. \ref{fig:log_neg_closed}(b). 

\begin{figure}
    \centering    
    \includegraphics[width=\linewidth]{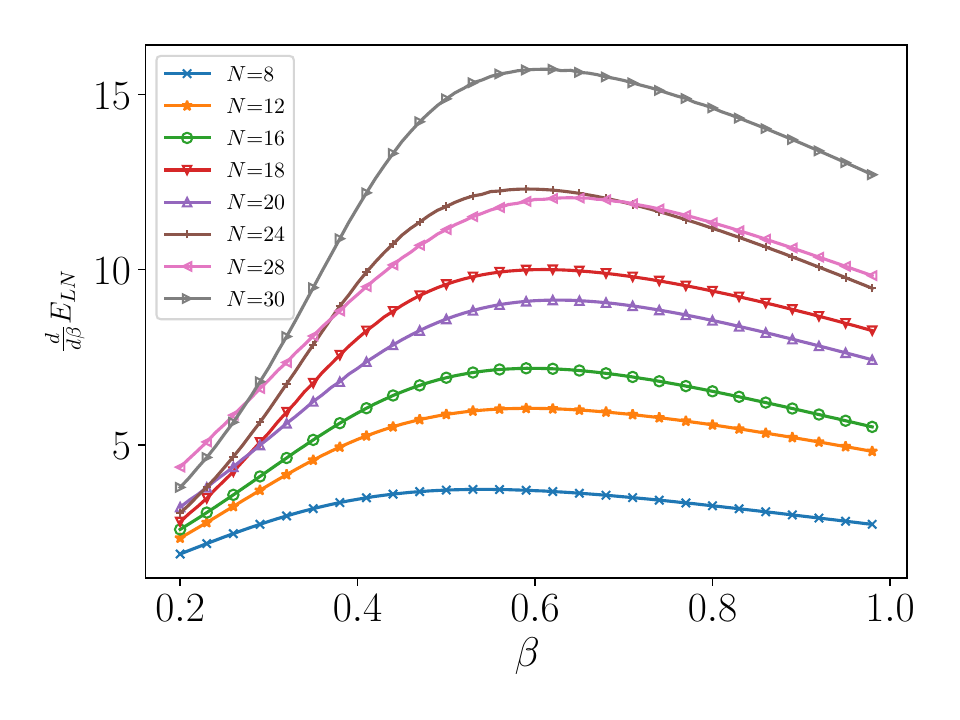}
    \caption{ The first derivative of block entanglement, $  \frac{d}{d \beta}E_{LN}\equiv \frac{d}{d \beta}E_{LN}(|GS\rangle)_{\frac{N}{2}:\frac{N}{2}}$), of the ground state (vertical axis) as a function of $\beta$ (horizontal axis). Note that, interestingly, it does not show a clear peak at $\beta$  close to the critical point as shown in the case of the dynamical state in Fig. \ref{fig:log_neg_closed}. Both axes are dimensionless.}
    \label{fig:log_neg_gs}
\end{figure}

\textit{Remark.} The block entanglement of the ground state in the $\frac{N}{2}:\frac{N}{2}$ bipartition does not change curvature with $\beta$, and hence, no divergence is observed in its derivative with respect to \(\beta\) (see Fig. \ref{fig:log_neg_gs}).  Interestingly, however, the time-averaged entanglement can predict the TQPT, as shown in Fig. \ref{fig:log_neg_closed}.

\begin{figure*}
    \centering
    \includegraphics[width=\linewidth]{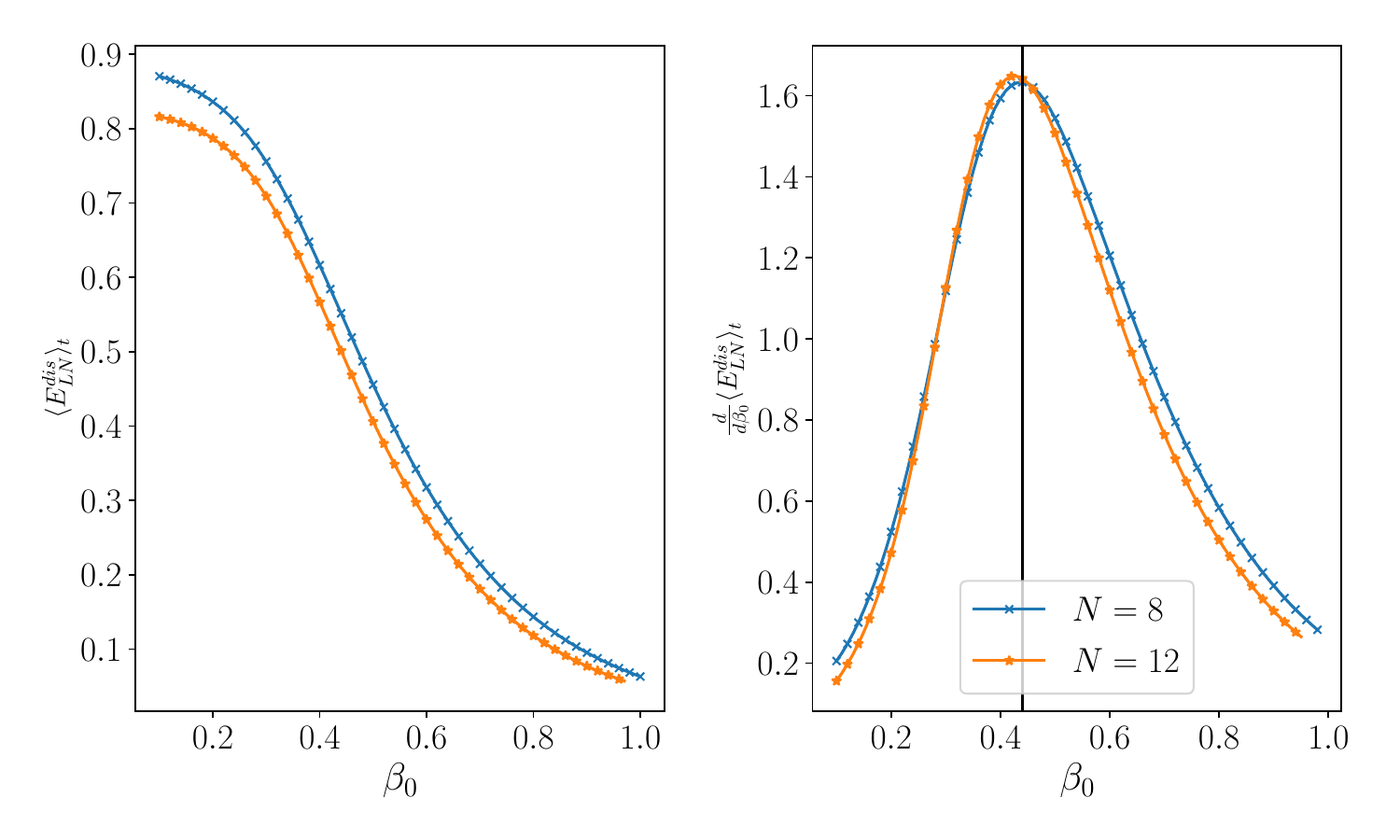}
    \caption{ Topological dynamical quantum phase transition under decoherence. The individual qubits of the entire system are in contact with a local thermal bath with temperature, \(B/T_E = 10\). 
    (a) Plot of the time-averaged LN in a repetitive noisy environment, $\langle E_{LN}^{dis} \rangle_t$ (ordinate) while  varying $\beta_0$ (abscissa) of a perturbed Kitaev toric code for different \(N\). (b) The first derivative of $\langle E_{LN}^{dis} \rangle_t$ (vertical axis) again shows a clear peak at $\beta_0 \sim \beta_c$ (horizontal axis),  close to the critical point. It indicates that even in the presence of 
    decoherence, the system carries information about the phase in the initial Hamiltonian. 
    Both axes are dimensionless. }
    \label{fig:log_neg_open}
\end{figure*}

\section{Effect of local repetitive interaction on the toric code}
\label{sec:criticaldecoh}

Until now, we have studied the unitary dynamics which means that the system is isolated and does not interact with the environment. We have shown that entanglement as well as the Loschmidt echo from the evolved state can faithfully signal the topological quantum phase transition present at zero temperature.  

Let us now ask whether the above measures of entanglement can identify the topological critical point from the dynamics of the system even when the system is in contact with a bath. It is quite reasonable to expect decoherence to erase the information about the equilibrium phase transition  from the evolved state. Below, we will show that this is not the case at least for the local repetitive interaction between the system and the bath. 



Before presenting the results, let us first describe the decoherence model considered here. We assume that the interaction between the system $S$ and the environment $E$ is repetitive in nature. The environment can be modeled by a thermal bath containing photons whose spin degree of freedom can be represented as spin-$1/2$ particles \cite{dhahri2008lindblad, Titas18}. The individual spin present in the environment interacts with a spin in the system for a short period of time, $\delta t$, and then a new spin from the environment interacts with the spin of the system for the next $\delta t$ time period and so on. The nature of this kind of interaction is repetitive, hence the name.  Further, it ensures Markovianity since the state of the system and the environment decouple at every $\delta t$ time-interval i.e., $\rho_{SE} = \rho_S \otimes \rho_E$ where $\rho_S$ gets modified at each $\delta t$ time-period as $\rho_S = \text{Tr}_E (U(\delta t) \rho_S \otimes \rho_E U^\dagger(\delta t))$ with $U(\delta t) $ being the unitary evolution acting on both the system and environment. Here, $\rho_E$ is taken to be the thermal state of a single qubit which always remains the same at every time-step. To facilitate such an interaction, the corresponding Hamiltonian between the spin and the environment is given by 
the $H_{SE} = H_S \otimes I_E + I_S \otimes H_E + H_{int}$, where $H_{int} = \sqrt{k/\delta t} (\hat{\sigma}^x_S \otimes \hat{\sigma}^x_E + \hat{\sigma}^y_S \otimes \hat{\sigma}^y_E)$. In our case, $H_S = H_{NLTC}$ as described in Eq. (\ref{H})  and $H_{E} = B \hat{\sigma}^z_i$, where $B$ is the strength of the local magnetic field acting on the bath. The parameter $k$ corresponds to the strength of the interaction between the bath and the system. The evolution of the system is governed by the  Gorini–Kossakowski–Sudarshan–Lindblad (GKSL) master equation \cite{Rivas_2014, noise1, noisebook}, which is written under the Born-Markov approximation  as 
\begin{equation}
\frac{d \rho_{S}}{d t}=-\frac{i}{\hbar}\left[H_{NLTC}, \rho_{S}\right]+\mathcal{D}\left(\rho_{S}\right),
\label{lindblad}
\end{equation}
where $\mathcal{D}$ is the dissipative part dictated by the choice of the environment.
The dissipative term in the Markovian limit with the assumption of weak coupling (i.e., we assume that interaction strength is much weaker than the local terms) reduces to 
 \begin{equation}
\mathcal{D}\left(\rho_{S}\right)=\frac{2 k}{\hbar^{2}}\sum_{i=1}^{N} \sum_{l=0}^{1} p_{l}\left[2 \eta_{i}^{l+1} \rho_{S} \eta_{i}^{l}-\left\{\eta_{i}^{l} \eta_{i}^{l+1}, \rho_{S}\right\}\right],
\label{repetative}
\end{equation}
where $p_{l}=Z_{E}^{-1} \exp \left[\frac{(-1)^{l}  B}{T_{E}}\right]$; $Z_{E}=\operatorname{tr}\left[\exp \left(\frac{-H_{E}}{T_{E}}\right)\right],$ 
 with the temperature of the bath Hamiltonian being \(T_E\) and $\eta_{i}^{\alpha}=(\hat{\sigma}_{i}^{x}+i(-1)^{\alpha} \hat{\sigma}_{i}^{y})/2$, with the subscript \(i\) denoting the spins which interact with the environment. Notice that the evolution contains the 
 Schr\"odinger part involving $H_{NLTC}$ and the dissipative term $\mathcal{D}$ as described above. Although the Lindblad operators act locally, the evolution is nonlocal due to the four-body operators present in the Hamiltonian [see Eq.~(\ref{H})], thereby leading to non-trivial evolution. Furthermore, in the presence of local noise, the entanglement decreases with the increase of noise and hence there are possibilities that noise can erase the information of the TQPT encoded in the unitary dynamics of entanglement. We will illustrate that this is not the case even though entanglement decays.

The initial state is again chosen to be the ground state of  $H_{NLTC}$ with different $\beta_0 \neq 0$. By solving the above master equation, we obtain \(\rho(t)\) at a given time which is used to compute the time-averaged LN by partitioning the entire system into equal blocks. Notice that due to the dissipative term, the evolved state is no longer a pure state, but a noisy density matrix, and hence, the computation of the GGM via Schmidt coefficients is not possible (see \cite{ggm_mixed_otfried, aditi_ggm_mixed}), although LN in the bipartition can be obtained.

For a fixed system size \(N\) and fixed temperature of the bath (for Fig. \ref{fig:log_neg_open}, \(B/T_E = 10\)),  we observe that  \(\langle E_{LN}^{dis}\rangle_t\) decreases substantially since all the qubits in the perturbed Kitaev toric code interact with the thermal bath repeatedly. However, the overall behavior of the time-averaged entanglement remains the same which can be seen by comparing Figs. \ref{fig:log_neg_open} and \ref{fig:log_neg_closed}. In particular, we find that when the initial state is prepared in the topologically ordered phase, the dynamical state even in the presence of a noisy environment can create high entanglement compared to the initial state prepared in the paramagnetic phase.  

Moreover, depending on the choice of the phase of the initial state, the curvature of \(\langle E_{LN}^{dis} (\rho)\rangle_t \equiv \langle E_{LN}^{dis} (\rho_{\frac{N}{2}:\frac{N}{2} })\rangle_t\) changes to convex at the TQPT point from concave and, therefore,  \(\frac{d\langle E_{LN}^{dis}\rangle_t}{dt}\) shows an increase at the quantum critical point which is in good agreement with the exact topological phase transition point. 


\section{Summary}
\label{sec:conclu}

In many-body systems, local order parameters are commonly employed to characterize quantum phases.  The topological order of the system, which is resilient to local disruption, is an exception. It is known that the nonlinearly perturbed Kitaev toric code undergoes a quantum phase transition from the topological phase to a paramagnetic one as the nonlinear perturbation parameter is tuned beyond a critical value. The best feature of this model is that the ground state can be found analytically and it also exhibits topological order which is robust against local perturbation. On the other hand, there is a constant effort to detect the QPT occurring at zero temperature by investigating the evolved state via sudden quench in distinct phases across the quantum critical point -- the phenomena is known as the dynamical quantum phase transition.  

In this context, we studied the quench dynamics of the nonlinearly perturbed Kitaev toric code and searched for the signatures of topological quantum phase transitions in the dynamics which we refer to as the topological dynamical quantum phase transition. We choose the initial state for different values of the perturbation
from both sides of the quantum critical point and quenched the system with the original Kitaev code. 
In addition to the conventional markers of a quantum critical point in the dynamical state, such as the Loschmidt echo and the rate function, we used various entanglement quantifiers, namely time-averaged genuine multipartite entanglement quantified via the generalized geometric measure, and block entanglement to detect the TDQPT. 
We found that under closed unitary dynamics, all the identifiers can distinguish whether the initial state is in the topological phase or the paramagnetic phase, thereby signaling the topological QPT via dynamics. Further, we observed that if both the initial and  final states belong to the topologically ordered phase, the average multipartite entanglement generated is comparatively higher than in the scenario in which they are chosen from different phases. Moreover, we reported that the block entanglement in the dynamical state can still identify the topological critical point when all individual qubits of the model interact with a thermal bath repeatedly. The results demonstrate that the topological quantum phase transition is prominent enough to be revealed in the dynamics with or without a noisy environment. 

\section*{acknowledgements}

We acknowledge the support from the Interdisciplinary Cyber Physical Systems (ICPS) program of the Department of Science and Technology (DST), India, Grant No.: DST/ICPS/QuST/Theme- 1/2019/23 and TARE Grant No.: TAR/2021/000136. We acknowledge the use of \href{https://github.com/titaschanda/QIClib}{QIClib} -- a modern C++ library for general-purpose quantum information processing and quantum computing (\url{https://titaschanda.github.io/QIClib}) and the high-performance computing facility at the Harish-Chandra Research Institute. 



\appendix
\section{Map to 2D Ising model and the topological quantum critical point}
\label{appendix:deri}

Let us briefly outline the derivation of the topological quantum phase transition point at zero temperature for the perturbed Kitaev toric code Hamiltonian \cite{Claduio2008, Zarei2022}. 
To do so, we provide a concise overview of the quantum formalism for the partition function of the two-dimensional (2D) Ising model. The model involves classical spins ($\tau = \pm 1$) residing on the vertices of a 2D square lattice, where the classical  Ising Hamiltonian is given by
\begin{equation*}
    H = -\sum_{\langle i, j\rangle} \tau_i \tau_j,
\end{equation*} 
where $\langle i, j\rangle$ denotes the interactions between nearest-neighbor spins.

The partition function, which plays a pivotal role in determining the thermodynamic properties of the model, is given by 
\begin{equation*}
    \mathcal{Z} = \sum_{\tau \in \Theta} e^{\beta \sum_{\la i, j\ra} \tau_i \tau_j},
\end{equation*}
where $\beta = \frac{1}{k_B T}$, with $k_B$ being the Boltzmann constant and $T$ being the equilibrium temperature. Here, $\Theta$ represents the set of all possible spin configurations.
Interestingly, the partition function can be reformulated using a quantum formalism. By introducing quantum spins ($S_e$) on the edges of each square grid of the lattice, each represented by the product of the vertex spins at the endpoints of the edge, i.e., $\la S_e|\hat{\sigma}^z|S_e \ra = \tau_i \tau_j$. This means that if the quantum spin is $|1\rangle $, one of the spins on either of the vertices in the Ising lattice must be $-1$ and the other one is $+1$. To represent the loops of the toric code, the following map has to be realized. Every loop in the toric code is generated via the star operator $A_\nu$, i.e., one element $a$ in the Abelian group $G$ acting on the vacuum $|0\rangle ^{\otimes N}$ with \(N\) being the total number of sites. In terms of the  Ising spins, if one spin on a vertex is $-1$ and the rest of the spins surrounding that spin take a value of $+1$, the corresponding edge spins of the toric code form a loop. Thus, for every $a \in G$,  two Ising spin configurations exist such that 
\begin{align}
 \sum_i \hat{\sigma}_i^z(a) = \la 0^{\otimes N} |a \left(\sum_{i=1}^N \hat{\sigma}^z_i \right) a | 0^{\otimes N} \ra\equiv \frac{1}{2} \sum_{\left\langle i, j \right\rangle} \tau_i \tau_j. 
\end{align}
Among several observables corresponding to the toric code that can be calculated in terms of the partition function of the 2D Ising model, the local magnetization can be written as
\begin{align}
m(\beta) & =\frac{1}{N} \sum_i\left\langle\mathrm{GS} (\beta) \left|\hat{\sigma}_i^z\right| \mathrm{GS}(\beta)\right\rangle \nonumber \\ 
& =\frac{1}{\mathcal{Z}} \sum_{a \in G} \exp \left(\beta \sum_i \hat{\sigma}_i^z(a)\right)\left[\frac{1}{N} \sum_i \hat{\sigma}_i^z(a)\right] \nonumber \\
& =\frac{1}{\mathcal{Z}} \sum_{\tau \in T} \exp \left( \beta \sum_{\left\langle i, j\right\rangle} \tau_i \tau_j \right)\left[\frac{1}{N} \sum_{\left\langle i, j\right\rangle} \tau_i \tau_j \right] \nonumber\\
& =\frac{1}{N} \text{Energy}_{\mathrm{Ising}}(\beta) .
\end{align}

We know that the 2D classical Ising model undergoes a first-order phase transition with the variation of the inverse temperature $\beta$. This can be seen from the fact that the first derivative of the energy of the model represents specific heat, which has non-analytic behavior around $\beta_c = \frac{\ln(\sqrt{2}+1)}{2} = 0.44068$ \cite{REIF65}. This also shows that the magnetization of the toric code changes curvature around the $\beta = \beta_c$ point, indicating a quantum phase transition point. By investigating the magnetic properties with $\beta >> \beta_c$, we observe that it always remains in the paramagnetic phase, thereby confirming that there is no further quantum phase transition with \(\beta\). Moreover, the transition is topological in nature, which can be confirmed by studying the topological entropy \cite{Claduio2008}. 


\bibliography{bib_topo}

\begin{thebibliography}{82}%
\makeatletter
\providecommand \@ifxundefined [1]{%
 \@ifx{#1\undefined}
}%
\providecommand \@ifnum [1]{%
 \ifnum #1\expandafter \@firstoftwo
 \else \expandafter \@secondoftwo
 \fi
}%
\providecommand \@ifx [1]{%
 \ifx #1\expandafter \@firstoftwo
 \else \expandafter \@secondoftwo
 \fi
}%
\providecommand \natexlab [1]{#1}%
\providecommand \enquote  [1]{``#1''}%
\providecommand \bibnamefont  [1]{#1}%
\providecommand \bibfnamefont [1]{#1}%
\providecommand \citenamefont [1]{#1}%
\providecommand \href@noop [0]{\@secondoftwo}%
\providecommand \href [0]{\begingroup \@sanitize@url \@href}%
\providecommand \@href[1]{\@@startlink{#1}\@@href}%
\providecommand \@@href[1]{\endgroup#1\@@endlink}%
\providecommand \@sanitize@url [0]{\catcode `\\12\catcode `\$12\catcode
  `\&12\catcode `\#12\catcode `\^12\catcode `\_12\catcode `\%12\relax}%
\providecommand \@@startlink[1]{}%
\providecommand \@@endlink[0]{}%
\providecommand \url  [0]{\begingroup\@sanitize@url \@url }%
\providecommand \@url [1]{\endgroup\@href {#1}{\urlprefix }}%
\providecommand \urlprefix  [0]{URL }%
\providecommand \Eprint [0]{\href }%
\providecommand \doibase [0]{https://doi.org/}%
\providecommand \selectlanguage [0]{\@gobble}%
\providecommand \bibinfo  [0]{\@secondoftwo}%
\providecommand \bibfield  [0]{\@secondoftwo}%
\providecommand \translation [1]{[#1]}%
\providecommand \BibitemOpen [0]{}%
\providecommand \bibitemStop [0]{}%
\providecommand \bibitemNoStop [0]{.\EOS\space}%
\providecommand \EOS [0]{\spacefactor3000\relax}%
\providecommand \BibitemShut  [1]{\csname bibitem#1\endcsname}%
\let\auto@bib@innerbib\@empty
\bibitem [{\citenamefont {Sachdev}(2011)}]{sachdev_2011}%
  \BibitemOpen
  \bibfield  {author} {\bibinfo {author} {\bibfnamefont {S.}~\bibnamefont
  {Sachdev}},\ }\href {https://doi.org/10.1017/CBO9780511973765} {\emph
  {\bibinfo {title} {Quantum Phase Transitions}}},\ \bibinfo {edition} {2nd}\
  ed.\ (\bibinfo  {publisher} {Cambridge University Press},\ \bibinfo {year}
  {2011})\BibitemShut {NoStop}%
\bibitem [{\citenamefont {Sengupta}\ \emph {et~al.}(2004)\citenamefont
  {Sengupta}, \citenamefont {Powell},\ and\ \citenamefont
  {Sachdev}}]{Krish2004}%
  \BibitemOpen
  \bibfield  {author} {\bibinfo {author} {\bibfnamefont {K.}~\bibnamefont
  {Sengupta}}, \bibinfo {author} {\bibfnamefont {S.}~\bibnamefont {Powell}},\
  and\ \bibinfo {author} {\bibfnamefont {S.}~\bibnamefont {Sachdev}},\ }\href
  {https://doi.org/10.1103/PhysRevA.69.053616} {\bibfield  {journal} {\bibinfo
  {journal} {Phys. Rev. A}\ }\textbf {\bibinfo {volume} {69}},\ \bibinfo
  {pages} {053616} (\bibinfo {year} {2004})}\BibitemShut {NoStop}%
\bibitem [{\citenamefont {Sen(De)}\ \emph {et~al.}(2005)\citenamefont
  {Sen(De)}, \citenamefont {Sen},\ and\ \citenamefont
  {Lewenstein}}]{Aditi2005}%
  \BibitemOpen
  \bibfield  {author} {\bibinfo {author} {\bibfnamefont {A.}~\bibnamefont
  {Sen(De)}}, \bibinfo {author} {\bibfnamefont {U.}~\bibnamefont {Sen}},\ and\
  \bibinfo {author} {\bibfnamefont {M.}~\bibnamefont {Lewenstein}},\ }\href
  {https://doi.org/10.1103/PhysRevA.72.052319} {\bibfield  {journal} {\bibinfo
  {journal} {Phys. Rev. A}\ }\textbf {\bibinfo {volume} {72}},\ \bibinfo
  {pages} {052319} (\bibinfo {year} {2005})}\BibitemShut {NoStop}%
\bibitem [{\citenamefont {Heyl}(2018)}]{Heylrev}%
  \BibitemOpen
  \bibfield  {author} {\bibinfo {author} {\bibfnamefont {M.}~\bibnamefont
  {Heyl}},\ }\href {https://doi.org/10.1088/1361-6633/aaaf9a} {\bibfield
  {journal} {\bibinfo  {journal} {Reports on Progress in Physics}\ }\textbf
  {\bibinfo {volume} {81}},\ \bibinfo {pages} {054001} (\bibinfo {year}
  {2018})}\BibitemShut {NoStop}%
\bibitem [{\citenamefont {Heyl}\ \emph {et~al.}(2013)\citenamefont {Heyl},
  \citenamefont {Polkovnikov},\ and\ \citenamefont {Kehrein}}]{heyl_original}%
  \BibitemOpen
  \bibfield  {author} {\bibinfo {author} {\bibfnamefont {M.}~\bibnamefont
  {Heyl}}, \bibinfo {author} {\bibfnamefont {A.}~\bibnamefont {Polkovnikov}},\
  and\ \bibinfo {author} {\bibfnamefont {S.}~\bibnamefont {Kehrein}},\ }\href
  {https://doi.org/10.1103/PhysRevLett.110.135704} {\bibfield  {journal}
  {\bibinfo  {journal} {Phys. Rev. Lett.}\ }\textbf {\bibinfo {volume} {110}},\
  \bibinfo {pages} {135704} (\bibinfo {year} {2013})}\BibitemShut {NoStop}%
\bibitem [{\citenamefont {Haldar}\ \emph {et~al.}(2021)\citenamefont {Haldar},
  \citenamefont {Mallayya}, \citenamefont {Heyl}, \citenamefont {Pollmann},
  \citenamefont {Rigol},\ and\ \citenamefont {Das}}]{arnab_dqpt_critical1}%
  \BibitemOpen
  \bibfield  {author} {\bibinfo {author} {\bibfnamefont {A.}~\bibnamefont
  {Haldar}}, \bibinfo {author} {\bibfnamefont {K.}~\bibnamefont {Mallayya}},
  \bibinfo {author} {\bibfnamefont {M.}~\bibnamefont {Heyl}}, \bibinfo {author}
  {\bibfnamefont {F.}~\bibnamefont {Pollmann}}, \bibinfo {author}
  {\bibfnamefont {M.}~\bibnamefont {Rigol}},\ and\ \bibinfo {author}
  {\bibfnamefont {A.}~\bibnamefont {Das}},\ }\href
  {https://doi.org/10.1103/PhysRevX.11.031062} {\bibfield  {journal} {\bibinfo
  {journal} {Phys. Rev. X}\ }\textbf {\bibinfo {volume} {11}},\ \bibinfo
  {pages} {031062} (\bibinfo {year} {2021})}\BibitemShut {NoStop}%
\bibitem [{\citenamefont {Bhattacharyya}\ \emph {et~al.}(2015)\citenamefont
  {Bhattacharyya}, \citenamefont {Dasgupta},\ and\ \citenamefont
  {Das}}]{arnab_dqpt_critical3}%
  \BibitemOpen
  \bibfield  {author} {\bibinfo {author} {\bibfnamefont {S.}~\bibnamefont
  {Bhattacharyya}}, \bibinfo {author} {\bibfnamefont {S.}~\bibnamefont
  {Dasgupta}},\ and\ \bibinfo {author} {\bibfnamefont {A.}~\bibnamefont
  {Das}},\ }\href {https://doi.org/10.1038/srep16490} {\bibfield  {journal}
  {\bibinfo  {journal} {Scientific Reports}\ }\textbf {\bibinfo {volume} {5}},\
  \bibinfo {pages} {16490} (\bibinfo {year} {2015})}\BibitemShut {NoStop}%
\bibitem [{\citenamefont {Jafari}\ \emph {et~al.}(2019)\citenamefont {Jafari},
  \citenamefont {Johannesson}, \citenamefont {Langari},\ and\ \citenamefont
  {Martin-Delgado}}]{jafari_dqpt2}%
  \BibitemOpen
  \bibfield  {author} {\bibinfo {author} {\bibfnamefont {R.}~\bibnamefont
  {Jafari}}, \bibinfo {author} {\bibfnamefont {H.}~\bibnamefont {Johannesson}},
  \bibinfo {author} {\bibfnamefont {A.}~\bibnamefont {Langari}},\ and\ \bibinfo
  {author} {\bibfnamefont {M.~A.}\ \bibnamefont {Martin-Delgado}},\ }\href
  {https://doi.org/10.1103/PhysRevB.99.054302} {\bibfield  {journal} {\bibinfo
  {journal} {Phys. Rev. B}\ }\textbf {\bibinfo {volume} {99}},\ \bibinfo
  {pages} {054302} (\bibinfo {year} {2019})}\BibitemShut {NoStop}%
\bibitem [{\citenamefont {Horodecki}\ \emph {et~al.}(2009)\citenamefont
  {Horodecki}, \citenamefont {Horodecki}, \citenamefont {Horodecki},\ and\
  \citenamefont {Horodecki}}]{HoroRMP}%
  \BibitemOpen
  \bibfield  {author} {\bibinfo {author} {\bibfnamefont {R.}~\bibnamefont
  {Horodecki}}, \bibinfo {author} {\bibfnamefont {P.}~\bibnamefont
  {Horodecki}}, \bibinfo {author} {\bibfnamefont {M.}~\bibnamefont
  {Horodecki}},\ and\ \bibinfo {author} {\bibfnamefont {K.}~\bibnamefont
  {Horodecki}},\ }\href {https://doi.org/10.1103/RevModPhys.81.865} {\bibfield
  {journal} {\bibinfo  {journal} {Rev. Mod. Phys.}\ }\textbf {\bibinfo {volume}
  {81}},\ \bibinfo {pages} {865} (\bibinfo {year} {2009})}\BibitemShut
  {NoStop}%
\bibitem [{\citenamefont {Wei}\ \emph {et~al.}(2005)\citenamefont {Wei},
  \citenamefont {Das}, \citenamefont {Mukhopadyay}, \citenamefont
  {Vishveshwara},\ and\ \citenamefont {Goldbart}}]{Wei05}%
  \BibitemOpen
  \bibfield  {author} {\bibinfo {author} {\bibfnamefont {T.-C.}\ \bibnamefont
  {Wei}}, \bibinfo {author} {\bibfnamefont {D.}~\bibnamefont {Das}}, \bibinfo
  {author} {\bibfnamefont {S.}~\bibnamefont {Mukhopadyay}}, \bibinfo {author}
  {\bibfnamefont {S.}~\bibnamefont {Vishveshwara}},\ and\ \bibinfo {author}
  {\bibfnamefont {P.~M.}\ \bibnamefont {Goldbart}},\ }\href
  {https://doi.org/10.1103/PhysRevA.71.060305} {\bibfield  {journal} {\bibinfo
  {journal} {Phys. Rev. A}\ }\textbf {\bibinfo {volume} {71}},\ \bibinfo
  {pages} {060305} (\bibinfo {year} {2005})}\BibitemShut {NoStop}%
\bibitem [{\citenamefont {Biswas}\ \emph {et~al.}(2014)\citenamefont {Biswas},
  \citenamefont {Prabhu}, \citenamefont {Sen(De)},\ and\ \citenamefont
  {Sen}}]{Anindya14}%
  \BibitemOpen
  \bibfield  {author} {\bibinfo {author} {\bibfnamefont {A.}~\bibnamefont
  {Biswas}}, \bibinfo {author} {\bibfnamefont {R.}~\bibnamefont {Prabhu}},
  \bibinfo {author} {\bibfnamefont {A.}~\bibnamefont {Sen(De)}},\ and\ \bibinfo
  {author} {\bibfnamefont {U.}~\bibnamefont {Sen}},\ }\href
  {https://doi.org/10.1103/PhysRevA.90.032301} {\bibfield  {journal} {\bibinfo
  {journal} {Phys. Rev. A}\ }\textbf {\bibinfo {volume} {90}},\ \bibinfo
  {pages} {032301} (\bibinfo {year} {2014})}\BibitemShut {NoStop}%
\bibitem [{\citenamefont {Lewenstein}\ \emph {et~al.}(2007)\citenamefont
  {Lewenstein}, \citenamefont {Sanpera}, \citenamefont {Ahufinger},
  \citenamefont {Damski}, \citenamefont {Sen(De)},\ and\ \citenamefont
  {Sen}}]{ultracold-review}%
  \BibitemOpen
  \bibfield  {author} {\bibinfo {author} {\bibfnamefont {M.}~\bibnamefont
  {Lewenstein}}, \bibinfo {author} {\bibfnamefont {A.}~\bibnamefont {Sanpera}},
  \bibinfo {author} {\bibfnamefont {V.}~\bibnamefont {Ahufinger}}, \bibinfo
  {author} {\bibfnamefont {B.}~\bibnamefont {Damski}}, \bibinfo {author}
  {\bibfnamefont {A.}~\bibnamefont {Sen(De)}},\ and\ \bibinfo {author}
  {\bibfnamefont {U.}~\bibnamefont {Sen}},\ }\href
  {https://doi.org/10.1080/00018730701223200} {\bibfield  {journal} {\bibinfo
  {journal} {Advances in Physics}\ }\textbf {\bibinfo {volume} {56}},\ \bibinfo
  {pages} {243} (\bibinfo {year} {2007})}\BibitemShut {NoStop}%
\bibitem [{\citenamefont {Amico}\ \emph {et~al.}(2008)\citenamefont {Amico},
  \citenamefont {Fazio}, \citenamefont {Osterloh},\ and\ \citenamefont
  {Vedral}}]{fazio-rev}%
  \BibitemOpen
  \bibfield  {author} {\bibinfo {author} {\bibfnamefont {L.}~\bibnamefont
  {Amico}}, \bibinfo {author} {\bibfnamefont {R.}~\bibnamefont {Fazio}},
  \bibinfo {author} {\bibfnamefont {A.}~\bibnamefont {Osterloh}},\ and\
  \bibinfo {author} {\bibfnamefont {V.}~\bibnamefont {Vedral}},\ }\href
  {https://doi.org/10.1103/RevModPhys.80.517} {\bibfield  {journal} {\bibinfo
  {journal} {Rev. Mod. Phys.}\ }\textbf {\bibinfo {volume} {80}},\ \bibinfo
  {pages} {517} (\bibinfo {year} {2008})}\BibitemShut {NoStop}%
\bibitem [{\citenamefont {Haldar}\ \emph {et~al.}(2020)\citenamefont {Haldar},
  \citenamefont {Roy}, \citenamefont {Chanda}, \citenamefont {Sen(De)},\ and\
  \citenamefont {Sen}}]{stavPRB}%
  \BibitemOpen
  \bibfield  {author} {\bibinfo {author} {\bibfnamefont {S.}~\bibnamefont
  {Haldar}}, \bibinfo {author} {\bibfnamefont {S.}~\bibnamefont {Roy}},
  \bibinfo {author} {\bibfnamefont {T.}~\bibnamefont {Chanda}}, \bibinfo
  {author} {\bibfnamefont {A.}~\bibnamefont {Sen(De)}},\ and\ \bibinfo {author}
  {\bibfnamefont {U.}~\bibnamefont {Sen}},\ }\href
  {https://doi.org/10.1103/PhysRevB.101.224304} {\bibfield  {journal} {\bibinfo
   {journal} {Phys. Rev. B}\ }\textbf {\bibinfo {volume} {101}},\ \bibinfo
  {pages} {224304} (\bibinfo {year} {2020})}\BibitemShut {NoStop}%
\bibitem [{\citenamefont {Häffner}\ \emph {et~al.}(2008)\citenamefont
  {Häffner}, \citenamefont {Roos},\ and\ \citenamefont
  {Blatt}}]{HAFFNER2008155}%
  \BibitemOpen
  \bibfield  {author} {\bibinfo {author} {\bibfnamefont {H.}~\bibnamefont
  {Häffner}}, \bibinfo {author} {\bibfnamefont {C.}~\bibnamefont {Roos}},\
  and\ \bibinfo {author} {\bibfnamefont {R.}~\bibnamefont {Blatt}},\ }\href
  {https://doi.org/https://doi.org/10.1016/j.physrep.2008.09.003} {\bibfield
  {journal} {\bibinfo  {journal} {Physics Reports}\ }\textbf {\bibinfo {volume}
  {469}},\ \bibinfo {pages} {155} (\bibinfo {year} {2008})}\BibitemShut
  {NoStop}%
\bibitem [{\citenamefont {Duan}\ and\ \citenamefont {Monroe}(2010)}]{Duan10}%
  \BibitemOpen
  \bibfield  {author} {\bibinfo {author} {\bibfnamefont {L.-M.}\ \bibnamefont
  {Duan}}\ and\ \bibinfo {author} {\bibfnamefont {C.}~\bibnamefont {Monroe}},\
  }\href {https://doi.org/10.1103/RevModPhys.82.1209} {\bibfield  {journal}
  {\bibinfo  {journal} {Rev. Mod. Phys.}\ }\textbf {\bibinfo {volume} {82}},\
  \bibinfo {pages} {1209} (\bibinfo {year} {2010})}\BibitemShut {NoStop}%
\bibitem [{\citenamefont {Kaur}\ \emph {et~al.}(2021)\citenamefont {Kaur},
  \citenamefont {S\'epulcre}, \citenamefont {Roch}, \citenamefont {Snyman},
  \citenamefont {Florens},\ and\ \citenamefont {Bera}}]{Kaur21}%
  \BibitemOpen
  \bibfield  {author} {\bibinfo {author} {\bibfnamefont {K.}~\bibnamefont
  {Kaur}}, \bibinfo {author} {\bibfnamefont {T.}~\bibnamefont {S\'epulcre}},
  \bibinfo {author} {\bibfnamefont {N.}~\bibnamefont {Roch}}, \bibinfo {author}
  {\bibfnamefont {I.}~\bibnamefont {Snyman}}, \bibinfo {author} {\bibfnamefont
  {S.}~\bibnamefont {Florens}},\ and\ \bibinfo {author} {\bibfnamefont
  {S.}~\bibnamefont {Bera}},\ }\href
  {https://doi.org/10.1103/PhysRevLett.127.237702} {\bibfield  {journal}
  {\bibinfo  {journal} {Phys. Rev. Lett.}\ }\textbf {\bibinfo {volume} {127}},\
  \bibinfo {pages} {237702} (\bibinfo {year} {2021})}\BibitemShut {NoStop}%
\bibitem [{\citenamefont {Wen}(1995)}]{Wen95}%
  \BibitemOpen
  \bibfield  {author} {\bibinfo {author} {\bibfnamefont {X.-G.}\ \bibnamefont
  {Wen}},\ }\href {https://doi.org/10.1080/00018739500101566} {\bibfield
  {journal} {\bibinfo  {journal} {Advances in Physics}\ }\textbf {\bibinfo
  {volume} {44}},\ \bibinfo {pages} {405} (\bibinfo {year} {1995})},\ \Eprint
  {https://arxiv.org/abs/https://doi.org/10.1080/00018739500101566}
  {https://doi.org/10.1080/00018739500101566} \BibitemShut {NoStop}%
\bibitem [{\citenamefont {Wen}(2002)}]{Wen02}%
  \BibitemOpen
  \bibfield  {author} {\bibinfo {author} {\bibfnamefont {X.-G.}\ \bibnamefont
  {Wen}},\ }\href {https://doi.org/10.1103/PhysRevB.65.165113} {\bibfield
  {journal} {\bibinfo  {journal} {Phys. Rev. B}\ }\textbf {\bibinfo {volume}
  {65}},\ \bibinfo {pages} {165113} (\bibinfo {year} {2002})}\BibitemShut
  {NoStop}%
\bibitem [{\citenamefont {Hamma}\ and\ \citenamefont
  {Lidar}(2008)}]{Hamma2008}%
  \BibitemOpen
  \bibfield  {author} {\bibinfo {author} {\bibfnamefont {A.}~\bibnamefont
  {Hamma}}\ and\ \bibinfo {author} {\bibfnamefont {D.~A.}\ \bibnamefont
  {Lidar}},\ }\href {https://doi.org/10.1103/PhysRevLett.100.030502} {\bibfield
   {journal} {\bibinfo  {journal} {Phys. Rev. Lett.}\ }\textbf {\bibinfo
  {volume} {100}},\ \bibinfo {pages} {030502} (\bibinfo {year}
  {2008})}\BibitemShut {NoStop}%
\bibitem [{\citenamefont {Trebst}\ \emph {et~al.}(2007)\citenamefont {Trebst},
  \citenamefont {Werner}, \citenamefont {Troyer}, \citenamefont {Shtengel},\
  and\ \citenamefont {Nayak}}]{Trebst07}%
  \BibitemOpen
  \bibfield  {author} {\bibinfo {author} {\bibfnamefont {S.}~\bibnamefont
  {Trebst}}, \bibinfo {author} {\bibfnamefont {P.}~\bibnamefont {Werner}},
  \bibinfo {author} {\bibfnamefont {M.}~\bibnamefont {Troyer}}, \bibinfo
  {author} {\bibfnamefont {K.}~\bibnamefont {Shtengel}},\ and\ \bibinfo
  {author} {\bibfnamefont {C.}~\bibnamefont {Nayak}},\ }\href
  {https://doi.org/10.1103/PhysRevLett.98.070602} {\bibfield  {journal}
  {\bibinfo  {journal} {Phys. Rev. Lett.}\ }\textbf {\bibinfo {volume} {98}},\
  \bibinfo {pages} {070602} (\bibinfo {year} {2007})}\BibitemShut {NoStop}%
\bibitem [{\citenamefont {Hamma}\ \emph {et~al.}(2008)\citenamefont {Hamma},
  \citenamefont {Zhang}, \citenamefont {Haas},\ and\ \citenamefont
  {Lidar}}]{PRB77}%
  \BibitemOpen
  \bibfield  {author} {\bibinfo {author} {\bibfnamefont {A.}~\bibnamefont
  {Hamma}}, \bibinfo {author} {\bibfnamefont {W.}~\bibnamefont {Zhang}},
  \bibinfo {author} {\bibfnamefont {S.}~\bibnamefont {Haas}},\ and\ \bibinfo
  {author} {\bibfnamefont {D.~A.}\ \bibnamefont {Lidar}},\ }\href
  {https://doi.org/10.1103/PhysRevB.77.155111} {\bibfield  {journal} {\bibinfo
  {journal} {Phys. Rev. B}\ }\textbf {\bibinfo {volume} {77}},\ \bibinfo
  {pages} {155111} (\bibinfo {year} {2008})}\BibitemShut {NoStop}%
\bibitem [{\citenamefont {Dennis}\ \emph {et~al.}(2002)\citenamefont {Dennis},
  \citenamefont {Kitaev}, \citenamefont {Landahl},\ and\ \citenamefont
  {Preskill}}]{Kitaev02}%
  \BibitemOpen
  \bibfield  {author} {\bibinfo {author} {\bibfnamefont {E.}~\bibnamefont
  {Dennis}}, \bibinfo {author} {\bibfnamefont {A.}~\bibnamefont {Kitaev}},
  \bibinfo {author} {\bibfnamefont {A.}~\bibnamefont {Landahl}},\ and\ \bibinfo
  {author} {\bibfnamefont {J.}~\bibnamefont {Preskill}},\ }\href
  {https://doi.org/10.1063/1.1499754} {\bibfield  {journal} {\bibinfo
  {journal} {Journal of Mathematical Physics}\ }\textbf {\bibinfo {volume}
  {43}},\ \bibinfo {pages} {4452} (\bibinfo {year} {2002})},\ \Eprint
  {https://arxiv.org/abs/https://doi.org/10.1063/1.1499754}
  {https://doi.org/10.1063/1.1499754} \BibitemShut {NoStop}%
\bibitem [{\citenamefont {Kitaev}(2003)}]{KITAEV20032}%
  \BibitemOpen
  \bibfield  {author} {\bibinfo {author} {\bibfnamefont {A.}~\bibnamefont
  {Kitaev}},\ }\href
  {https://doi.org/https://doi.org/10.1016/S0003-4916(02)00018-0} {\bibfield
  {journal} {\bibinfo  {journal} {Annals of Physics}\ }\textbf {\bibinfo
  {volume} {303}},\ \bibinfo {pages} {2} (\bibinfo {year} {2003})}\BibitemShut
  {NoStop}%
\bibitem [{\citenamefont {Kitaev}(2006)}]{KITAEV20062}%
  \BibitemOpen
  \bibfield  {author} {\bibinfo {author} {\bibfnamefont {A.}~\bibnamefont
  {Kitaev}},\ }\href
  {https://doi.org/https://doi.org/10.1016/j.aop.2005.10.005} {\bibfield
  {journal} {\bibinfo  {journal} {Annals of Physics}\ }\textbf {\bibinfo
  {volume} {321}},\ \bibinfo {pages} {2} (\bibinfo {year} {2006})},\ \bibinfo
  {note} {january Special Issue}\BibitemShut {NoStop}%
\bibitem [{\citenamefont {Castelnovo}\ and\ \citenamefont
  {Chamon}(2008)}]{Claduio2008}%
  \BibitemOpen
  \bibfield  {author} {\bibinfo {author} {\bibfnamefont {C.}~\bibnamefont
  {Castelnovo}}\ and\ \bibinfo {author} {\bibfnamefont {C.}~\bibnamefont
  {Chamon}},\ }\href {https://doi.org/10.1103/PhysRevB.77.054433} {\bibfield
  {journal} {\bibinfo  {journal} {Phys. Rev. B}\ }\textbf {\bibinfo {volume}
  {77}},\ \bibinfo {pages} {054433} (\bibinfo {year} {2008})}\BibitemShut
  {NoStop}%
\bibitem [{\citenamefont {Wu}\ \emph {et~al.}(2012)\citenamefont {Wu},
  \citenamefont {Deng},\ and\ \citenamefont {Prokof'ev}}]{Wu12}%
  \BibitemOpen
  \bibfield  {author} {\bibinfo {author} {\bibfnamefont {F.}~\bibnamefont
  {Wu}}, \bibinfo {author} {\bibfnamefont {Y.}~\bibnamefont {Deng}},\ and\
  \bibinfo {author} {\bibfnamefont {N.}~\bibnamefont {Prokof'ev}},\ }\href
  {https://doi.org/10.1103/PhysRevB.85.195104} {\bibfield  {journal} {\bibinfo
  {journal} {Phys. Rev. B}\ }\textbf {\bibinfo {volume} {85}},\ \bibinfo
  {pages} {195104} (\bibinfo {year} {2012})}\BibitemShut {NoStop}%
\bibitem [{\citenamefont {Nayak}\ \emph {et~al.}(2008)\citenamefont {Nayak},
  \citenamefont {Simon}, \citenamefont {Stern}, \citenamefont {Freedman},\ and\
  \citenamefont {Das~Sarma}}]{RevModPhys.80.1083}%
  \BibitemOpen
  \bibfield  {author} {\bibinfo {author} {\bibfnamefont {C.}~\bibnamefont
  {Nayak}}, \bibinfo {author} {\bibfnamefont {S.~H.}\ \bibnamefont {Simon}},
  \bibinfo {author} {\bibfnamefont {A.}~\bibnamefont {Stern}}, \bibinfo
  {author} {\bibfnamefont {M.}~\bibnamefont {Freedman}},\ and\ \bibinfo
  {author} {\bibfnamefont {S.}~\bibnamefont {Das~Sarma}},\ }\href
  {https://doi.org/10.1103/RevModPhys.80.1083} {\bibfield  {journal} {\bibinfo
  {journal} {Rev. Mod. Phys.}\ }\textbf {\bibinfo {volume} {80}},\ \bibinfo
  {pages} {1083} (\bibinfo {year} {2008})}\BibitemShut {NoStop}%
\bibitem [{\citenamefont {Levin}\ and\ \citenamefont {Wen}(2006)}]{Levin2006}%
  \BibitemOpen
  \bibfield  {author} {\bibinfo {author} {\bibfnamefont {M.}~\bibnamefont
  {Levin}}\ and\ \bibinfo {author} {\bibfnamefont {X.~G.}\ \bibnamefont
  {Wen}},\ }\href
  {https://doi.org/https://doi.org/10.1103/PhysRevLett.96.110405} {\bibfield
  {journal} {\bibinfo  {journal} {Phys. Rev. Lett.}\ }\textbf {\bibinfo
  {volume} {96}},\ \bibinfo {pages} {110405} (\bibinfo {year}
  {2006})}\BibitemShut {NoStop}%
\bibitem [{\citenamefont {Kitaev}\ and\ \citenamefont
  {Preskill}(2006)}]{Kitaev2006}%
  \BibitemOpen
  \bibfield  {author} {\bibinfo {author} {\bibfnamefont {A.}~\bibnamefont
  {Kitaev}}\ and\ \bibinfo {author} {\bibfnamefont {J.}~\bibnamefont
  {Preskill}},\ }\href
  {https://doi.org/https://doi.org/10.1103/PhysRevLett.96.110404} {\bibfield
  {journal} {\bibinfo  {journal} {Phys. Rev. Lett.}\ }\textbf {\bibinfo
  {volume} {96}},\ \bibinfo {pages} {110404} (\bibinfo {year}
  {2006})}\BibitemShut {NoStop}%
\bibitem [{\citenamefont {Jiang}\ \emph {et~al.}(2012)\citenamefont {Jiang},
  \citenamefont {Wang},\ and\ \citenamefont {Balents}}]{Jiang2012}%
  \BibitemOpen
  \bibfield  {author} {\bibinfo {author} {\bibfnamefont {H.-C.}\ \bibnamefont
  {Jiang}}, \bibinfo {author} {\bibfnamefont {Z.}~\bibnamefont {Wang}},\ and\
  \bibinfo {author} {\bibfnamefont {L.}~\bibnamefont {Balents}},\ }\href
  {https://doi.org/10.1038/nphys2465} {\bibfield  {journal} {\bibinfo
  {journal} {Nature Physics}\ }\textbf {\bibinfo {volume} {8}},\ \bibinfo
  {pages} {902} (\bibinfo {year} {2012})}\BibitemShut {NoStop}%
\bibitem [{\citenamefont {Yao}\ and\ \citenamefont {Qi}(2010)}]{Yao2010}%
  \BibitemOpen
  \bibfield  {author} {\bibinfo {author} {\bibfnamefont {H.}~\bibnamefont
  {Yao}}\ and\ \bibinfo {author} {\bibfnamefont {X.-L.}\ \bibnamefont {Qi}},\
  }\href {https://doi.org/https://doi.org/10.1103/PhysRevLett.105.080501}
  {\bibfield  {journal} {\bibinfo  {journal} {Phys. Rev. Lett.}\ }\textbf
  {\bibinfo {volume} {105}},\ \bibinfo {pages} {080501} (\bibinfo {year}
  {2010})}\BibitemShut {NoStop}%
\bibitem [{\citenamefont {Abasto}\ \emph {et~al.}(2008)\citenamefont {Abasto},
  \citenamefont {Hamma},\ and\ \citenamefont {Zanardi}}]{Abasto2008}%
  \BibitemOpen
  \bibfield  {author} {\bibinfo {author} {\bibfnamefont {D.~F.}\ \bibnamefont
  {Abasto}}, \bibinfo {author} {\bibfnamefont {A.}~\bibnamefont {Hamma}},\ and\
  \bibinfo {author} {\bibfnamefont {P.}~\bibnamefont {Zanardi}},\ }\href
  {https://doi.org/10.1103/PhysRevA.78.010301} {\bibfield  {journal} {\bibinfo
  {journal} {Phys. Rev. A}\ }\textbf {\bibinfo {volume} {78}},\ \bibinfo
  {pages} {010301} (\bibinfo {year} {2008})}\BibitemShut {NoStop}%
\bibitem [{\citenamefont {Chen}\ and\ \citenamefont {Li}(2010)}]{Chen2010}%
  \BibitemOpen
  \bibfield  {author} {\bibinfo {author} {\bibfnamefont {Y.~X.}\ \bibnamefont
  {Chen}}\ and\ \bibinfo {author} {\bibfnamefont {S.~W.}\ \bibnamefont {Li}},\
  }\href@noop {} {\bibfield  {journal} {\bibinfo  {journal} {Phys. Rev. A}\
  }\textbf {\bibinfo {volume} {81}},\ \bibinfo {pages} {032120} (\bibinfo
  {year} {(2010)})}\BibitemShut {NoStop}%
\bibitem [{\citenamefont {Zhang}\ \emph {et~al.}(2022)\citenamefont {Zhang},
  \citenamefont {Zeng}, \citenamefont {Liu}, \citenamefont {Fan}, \citenamefont
  {You},\ and\ \citenamefont {Nori}}]{Zhang2022}%
  \BibitemOpen
  \bibfield  {author} {\bibinfo {author} {\bibfnamefont {Y.-R.}\ \bibnamefont
  {Zhang}}, \bibinfo {author} {\bibfnamefont {Y.}~\bibnamefont {Zeng}},
  \bibinfo {author} {\bibfnamefont {T.}~\bibnamefont {Liu}}, \bibinfo {author}
  {\bibfnamefont {H.}~\bibnamefont {Fan}}, \bibinfo {author} {\bibfnamefont
  {J.~Q.}\ \bibnamefont {You}},\ and\ \bibinfo {author} {\bibfnamefont
  {F.}~\bibnamefont {Nori}},\ }\href
  {https://doi.org/10.1103/PhysRevResearch.4.023144} {\bibfield  {journal}
  {\bibinfo  {journal} {Phys. Rev. Res.}\ }\textbf {\bibinfo {volume} {4}},\
  \bibinfo {pages} {023144} (\bibinfo {year} {2022})}\BibitemShut {NoStop}%
\bibitem [{\citenamefont {Schotte}\ \emph {et~al.}(2019)\citenamefont
  {Schotte}, \citenamefont {Carrasco}, \citenamefont {Vanhecke}, \citenamefont
  {Vanderstraeten}, \citenamefont {Haegeman}, \citenamefont {Verstraete},\ and\
  \citenamefont {Vidal}}]{Schotte2019}%
  \BibitemOpen
  \bibfield  {author} {\bibinfo {author} {\bibfnamefont {A.}~\bibnamefont
  {Schotte}}, \bibinfo {author} {\bibfnamefont {J.}~\bibnamefont {Carrasco}},
  \bibinfo {author} {\bibfnamefont {B.}~\bibnamefont {Vanhecke}}, \bibinfo
  {author} {\bibfnamefont {L.}~\bibnamefont {Vanderstraeten}}, \bibinfo
  {author} {\bibfnamefont {J.}~\bibnamefont {Haegeman}}, \bibinfo {author}
  {\bibfnamefont {F.}~\bibnamefont {Verstraete}},\ and\ \bibinfo {author}
  {\bibfnamefont {J.}~\bibnamefont {Vidal}},\ }\href
  {https://doi.org/10.1103/PhysRevB.100.245125} {\bibfield  {journal} {\bibinfo
   {journal} {Phys. Rev. B}\ }\textbf {\bibinfo {volume} {100}},\ \bibinfo
  {pages} {245125} (\bibinfo {year} {2019})}\BibitemShut {NoStop}%
\bibitem [{\citenamefont {Chung}\ \emph {et~al.}(2010)\citenamefont {Chung},
  \citenamefont {Yao}, \citenamefont {Hughes},\ and\ \citenamefont
  {Kim}}]{entropy_tqpt2}%
  \BibitemOpen
  \bibfield  {author} {\bibinfo {author} {\bibfnamefont {S.~B.}\ \bibnamefont
  {Chung}}, \bibinfo {author} {\bibfnamefont {H.}~\bibnamefont {Yao}}, \bibinfo
  {author} {\bibfnamefont {T.~L.}\ \bibnamefont {Hughes}},\ and\ \bibinfo
  {author} {\bibfnamefont {E.-A.}\ \bibnamefont {Kim}},\ }\href
  {https://doi.org/10.1103/PhysRevB.81.060403} {\bibfield  {journal} {\bibinfo
  {journal} {Phys. Rev. B}\ }\textbf {\bibinfo {volume} {81}},\ \bibinfo
  {pages} {060403} (\bibinfo {year} {2010})}\BibitemShut {NoStop}%
\bibitem [{\citenamefont {Wang}\ \emph {et~al.}(2013)\citenamefont {Wang},
  \citenamefont {Li},\ and\ \citenamefont {Cho}}]{entropy_tqpt3}%
  \BibitemOpen
  \bibfield  {author} {\bibinfo {author} {\bibfnamefont {H.~T.}\ \bibnamefont
  {Wang}}, \bibinfo {author} {\bibfnamefont {B.}~\bibnamefont {Li}},\ and\
  \bibinfo {author} {\bibfnamefont {S.~Y.}\ \bibnamefont {Cho}},\ }\href
  {https://doi.org/10.1103/PhysRevB.87.054402} {\bibfield  {journal} {\bibinfo
  {journal} {Phys. Rev. B}\ }\textbf {\bibinfo {volume} {87}},\ \bibinfo
  {pages} {054402} (\bibinfo {year} {2013})}\BibitemShut {NoStop}%
\bibitem [{\citenamefont {Samimi}\ \emph {et~al.}(2022)\citenamefont {Samimi},
  \citenamefont {Zarei},\ and\ \citenamefont {Montakhab}}]{Zarei2022}%
  \BibitemOpen
  \bibfield  {author} {\bibinfo {author} {\bibfnamefont {E.}~\bibnamefont
  {Samimi}}, \bibinfo {author} {\bibfnamefont {M.~H.}\ \bibnamefont {Zarei}},\
  and\ \bibinfo {author} {\bibfnamefont {A.}~\bibnamefont {Montakhab}},\ }\href
  {https://doi.org/10.1103/PhysRevA.105.032438} {\bibfield  {journal} {\bibinfo
   {journal} {Phys. Rev. A}\ }\textbf {\bibinfo {volume} {105}},\ \bibinfo
  {pages} {032438} (\bibinfo {year} {2022})}\BibitemShut {NoStop}%
\bibitem [{\citenamefont {K.J.}\ and\ \citenamefont {Pal}(2022)}]{amit22}%
  \BibitemOpen
  \bibfield  {author} {\bibinfo {author} {\bibfnamefont {H.}~\bibnamefont
  {K.J.}}\ and\ \bibinfo {author} {\bibfnamefont {A.~K.}\ \bibnamefont {Pal}},\
  }\href {https://doi.org/10.1103/PhysRevA.105.052421} {\bibfield  {journal}
  {\bibinfo  {journal} {Phys. Rev. A}\ }\textbf {\bibinfo {volume} {105}},\
  \bibinfo {pages} {052421} (\bibinfo {year} {2022})}\BibitemShut {NoStop}%
\bibitem [{\citenamefont {Roy}\ \emph {et~al.}(2017)\citenamefont {Roy},
  \citenamefont {Moessner},\ and\ \citenamefont {Das}}]{arnab_tdqpt_critical2}%
  \BibitemOpen
  \bibfield  {author} {\bibinfo {author} {\bibfnamefont {S.}~\bibnamefont
  {Roy}}, \bibinfo {author} {\bibfnamefont {R.}~\bibnamefont {Moessner}},\ and\
  \bibinfo {author} {\bibfnamefont {A.}~\bibnamefont {Das}},\ }\href
  {https://doi.org/10.1103/PhysRevB.95.041105} {\bibfield  {journal} {\bibinfo
  {journal} {Phys. Rev. B}\ }\textbf {\bibinfo {volume} {95}},\ \bibinfo
  {pages} {041105} (\bibinfo {year} {2017})}\BibitemShut {NoStop}%
\bibitem [{\citenamefont {Uhrich}\ \emph {et~al.}(2020)\citenamefont {Uhrich},
  \citenamefont {Defenu}, \citenamefont {Jafari},\ and\ \citenamefont
  {Halimeh}}]{jafari_longrange_kitaev_dqpt2}%
  \BibitemOpen
  \bibfield  {author} {\bibinfo {author} {\bibfnamefont {P.}~\bibnamefont
  {Uhrich}}, \bibinfo {author} {\bibfnamefont {N.}~\bibnamefont {Defenu}},
  \bibinfo {author} {\bibfnamefont {R.}~\bibnamefont {Jafari}},\ and\ \bibinfo
  {author} {\bibfnamefont {J.~C.}\ \bibnamefont {Halimeh}},\ }\href
  {https://doi.org/10.1103/PhysRevB.101.245148} {\bibfield  {journal} {\bibinfo
   {journal} {Phys. Rev. B}\ }\textbf {\bibinfo {volume} {101}},\ \bibinfo
  {pages} {245148} (\bibinfo {year} {2020})}\BibitemShut {NoStop}%
\bibitem [{\citenamefont {Mishra}\ \emph {et~al.}(2020)\citenamefont {Mishra},
  \citenamefont {Jafari},\ and\ \citenamefont
  {Akbari}}]{jafari_utkarsh_longrange_kitaev_dqpt1}%
  \BibitemOpen
  \bibfield  {author} {\bibinfo {author} {\bibfnamefont {U.}~\bibnamefont
  {Mishra}}, \bibinfo {author} {\bibfnamefont {R.}~\bibnamefont {Jafari}},\
  and\ \bibinfo {author} {\bibfnamefont {A.}~\bibnamefont {Akbari}},\ }\href
  {https://doi.org/10.1088/1751-8121/ab97de} {\bibfield  {journal} {\bibinfo
  {journal} {Journal of Physics A: Mathematical and Theoretical}\ }\textbf
  {\bibinfo {volume} {53}},\ \bibinfo {pages} {375301} (\bibinfo {year}
  {2020})}\BibitemShut {NoStop}%
\bibitem [{\citenamefont {Naji}\ \emph
  {et~al.}(2022{\natexlab{a}})\citenamefont {Naji}, \citenamefont {Jafari},
  \citenamefont {Zhou},\ and\ \citenamefont {Langari}}]{jafari_floquet_dqpt1}%
  \BibitemOpen
  \bibfield  {author} {\bibinfo {author} {\bibfnamefont {J.}~\bibnamefont
  {Naji}}, \bibinfo {author} {\bibfnamefont {R.}~\bibnamefont {Jafari}},
  \bibinfo {author} {\bibfnamefont {L.}~\bibnamefont {Zhou}},\ and\ \bibinfo
  {author} {\bibfnamefont {A.}~\bibnamefont {Langari}},\ }\href
  {https://doi.org/10.1103/PhysRevB.106.094314} {\bibfield  {journal} {\bibinfo
   {journal} {Phys. Rev. B}\ }\textbf {\bibinfo {volume} {106}},\ \bibinfo
  {pages} {094314} (\bibinfo {year} {2022}{\natexlab{a}})}\BibitemShut
  {NoStop}%
\bibitem [{\citenamefont {Jafari}\ \emph {et~al.}(2022)\citenamefont {Jafari},
  \citenamefont {Akbari}, \citenamefont {Mishra},\ and\ \citenamefont
  {Johannesson}}]{jafari_floquet_dqpt2}%
  \BibitemOpen
  \bibfield  {author} {\bibinfo {author} {\bibfnamefont {R.}~\bibnamefont
  {Jafari}}, \bibinfo {author} {\bibfnamefont {A.}~\bibnamefont {Akbari}},
  \bibinfo {author} {\bibfnamefont {U.}~\bibnamefont {Mishra}},\ and\ \bibinfo
  {author} {\bibfnamefont {H.}~\bibnamefont {Johannesson}},\ }\href
  {https://doi.org/10.1103/PhysRevB.105.094311} {\bibfield  {journal} {\bibinfo
   {journal} {Phys. Rev. B}\ }\textbf {\bibinfo {volume} {105}},\ \bibinfo
  {pages} {094311} (\bibinfo {year} {2022})}\BibitemShut {NoStop}%
\bibitem [{\citenamefont {Zamani}\ \emph {et~al.}(2020)\citenamefont {Zamani},
  \citenamefont {Jafari},\ and\ \citenamefont
  {Langari}}]{jafari_floquet_dqpt3}%
  \BibitemOpen
  \bibfield  {author} {\bibinfo {author} {\bibfnamefont {S.}~\bibnamefont
  {Zamani}}, \bibinfo {author} {\bibfnamefont {R.}~\bibnamefont {Jafari}},\
  and\ \bibinfo {author} {\bibfnamefont {A.}~\bibnamefont {Langari}},\ }\href
  {https://doi.org/10.1103/PhysRevB.102.144306} {\bibfield  {journal} {\bibinfo
   {journal} {Phys. Rev. B}\ }\textbf {\bibinfo {volume} {102}},\ \bibinfo
  {pages} {144306} (\bibinfo {year} {2020})}\BibitemShut {NoStop}%
\bibitem [{\citenamefont {Jafari}\ and\ \citenamefont
  {Akbari}(2021)}]{jafari_floquet_dqpt4}%
  \BibitemOpen
  \bibfield  {author} {\bibinfo {author} {\bibfnamefont {R.}~\bibnamefont
  {Jafari}}\ and\ \bibinfo {author} {\bibfnamefont {A.}~\bibnamefont
  {Akbari}},\ }\href {https://doi.org/10.1103/PhysRevA.103.012204} {\bibfield
  {journal} {\bibinfo  {journal} {Phys. Rev. A}\ }\textbf {\bibinfo {volume}
  {103}},\ \bibinfo {pages} {012204} (\bibinfo {year} {2021})}\BibitemShut
  {NoStop}%
\bibitem [{\citenamefont {Sadrzadeh}\ \emph {et~al.}(2021)\citenamefont
  {Sadrzadeh}, \citenamefont {Jafari},\ and\ \citenamefont
  {Langari}}]{jafari_critical_dqpt}%
  \BibitemOpen
  \bibfield  {author} {\bibinfo {author} {\bibfnamefont {M.}~\bibnamefont
  {Sadrzadeh}}, \bibinfo {author} {\bibfnamefont {R.}~\bibnamefont {Jafari}},\
  and\ \bibinfo {author} {\bibfnamefont {A.}~\bibnamefont {Langari}},\ }\href
  {https://doi.org/10.1103/PhysRevB.103.144305} {\bibfield  {journal} {\bibinfo
   {journal} {Phys. Rev. B}\ }\textbf {\bibinfo {volume} {103}},\ \bibinfo
  {pages} {144305} (\bibinfo {year} {2021})}\BibitemShut {NoStop}%
\bibitem [{\citenamefont {Jangjan}\ \emph {et~al.}(2022)\citenamefont
  {Jangjan}, \citenamefont {Foa~Torres},\ and\ \citenamefont
  {Hosseini}}]{floquet_milad1}%
  \BibitemOpen
  \bibfield  {author} {\bibinfo {author} {\bibfnamefont {M.}~\bibnamefont
  {Jangjan}}, \bibinfo {author} {\bibfnamefont {L.~E.~F.}\ \bibnamefont
  {Foa~Torres}},\ and\ \bibinfo {author} {\bibfnamefont {M.~V.}\ \bibnamefont
  {Hosseini}},\ }\href {https://doi.org/10.1103/PhysRevB.106.224306} {\bibfield
   {journal} {\bibinfo  {journal} {Phys. Rev. B}\ }\textbf {\bibinfo {volume}
  {106}},\ \bibinfo {pages} {224306} (\bibinfo {year} {2022})}\BibitemShut
  {NoStop}%
\bibitem [{\citenamefont {Naji}\ \emph
  {et~al.}(2022{\natexlab{b}})\citenamefont {Naji}, \citenamefont {Jafari},
  \citenamefont {Jafari},\ and\ \citenamefont
  {Akbari}}]{jafari_floquet_nonhermitian_topological_dqpt}%
  \BibitemOpen
  \bibfield  {author} {\bibinfo {author} {\bibfnamefont {J.}~\bibnamefont
  {Naji}}, \bibinfo {author} {\bibfnamefont {M.}~\bibnamefont {Jafari}},
  \bibinfo {author} {\bibfnamefont {R.}~\bibnamefont {Jafari}},\ and\ \bibinfo
  {author} {\bibfnamefont {A.}~\bibnamefont {Akbari}},\ }\href
  {https://doi.org/10.1103/PhysRevA.105.022220} {\bibfield  {journal} {\bibinfo
   {journal} {Phys. Rev. A}\ }\textbf {\bibinfo {volume} {105}},\ \bibinfo
  {pages} {022220} (\bibinfo {year} {2022}{\natexlab{b}})}\BibitemShut
  {NoStop}%
\bibitem [{\citenamefont {Dusuel}\ \emph {et~al.}(2011)\citenamefont {Dusuel},
  \citenamefont {Kamfor}, \citenamefont {Or\'us}, \citenamefont {Schmidt},\
  and\ \citenamefont {Vidal}}]{Dusuel2011}%
  \BibitemOpen
  \bibfield  {author} {\bibinfo {author} {\bibfnamefont {S.}~\bibnamefont
  {Dusuel}}, \bibinfo {author} {\bibfnamefont {M.}~\bibnamefont {Kamfor}},
  \bibinfo {author} {\bibfnamefont {R.}~\bibnamefont {Or\'us}}, \bibinfo
  {author} {\bibfnamefont {K.~P.}\ \bibnamefont {Schmidt}},\ and\ \bibinfo
  {author} {\bibfnamefont {J.}~\bibnamefont {Vidal}},\ }\href
  {https://doi.org/10.1103/PhysRevLett.106.107203} {\bibfield  {journal}
  {\bibinfo  {journal} {Phys. Rev. Lett.}\ }\textbf {\bibinfo {volume} {106}},\
  \bibinfo {pages} {107203} (\bibinfo {year} {2011})}\BibitemShut {NoStop}%
\bibitem [{\citenamefont {Zarei}(2015)}]{Zarei2015}%
  \BibitemOpen
  \bibfield  {author} {\bibinfo {author} {\bibfnamefont {M.~H.}\ \bibnamefont
  {Zarei}},\ }\href {https://doi.org/10.1103/PhysRevA.91.022319} {\bibfield
  {journal} {\bibinfo  {journal} {Phys. Rev. A}\ }\textbf {\bibinfo {volume}
  {91}},\ \bibinfo {pages} {022319} (\bibinfo {year} {2015})}\BibitemShut
  {NoStop}%
\bibitem [{\citenamefont {Zarei}(2019)}]{Zarei2019}%
  \BibitemOpen
  \bibfield  {author} {\bibinfo {author} {\bibfnamefont {M.~H.}\ \bibnamefont
  {Zarei}},\ }\href {https://doi.org/10.1103/PhysRevB.100.125159} {\bibfield
  {journal} {\bibinfo  {journal} {Phys. Rev. B}\ }\textbf {\bibinfo {volume}
  {100}},\ \bibinfo {pages} {125159} (\bibinfo {year} {2019})}\BibitemShut
  {NoStop}%
\bibitem [{\citenamefont {Vidal}\ \emph {et~al.}(2009)\citenamefont {Vidal},
  \citenamefont {Thomale}, \citenamefont {Schmidt},\ and\ \citenamefont
  {Dusuel}}]{Vidal2009}%
  \BibitemOpen
  \bibfield  {author} {\bibinfo {author} {\bibfnamefont {J.}~\bibnamefont
  {Vidal}}, \bibinfo {author} {\bibfnamefont {R.}~\bibnamefont {Thomale}},
  \bibinfo {author} {\bibfnamefont {K.~P.}\ \bibnamefont {Schmidt}},\ and\
  \bibinfo {author} {\bibfnamefont {S.}~\bibnamefont {Dusuel}},\ }\href
  {https://doi.org/10.1103/PhysRevB.80.081104} {\bibfield  {journal} {\bibinfo
  {journal} {Phys. Rev. B}\ }\textbf {\bibinfo {volume} {80}},\ \bibinfo
  {pages} {081104} (\bibinfo {year} {2009})}\BibitemShut {NoStop}%
\bibitem [{\citenamefont {Sen(De)}\ and\ \citenamefont
  {Sen}(2010)}]{ggm_aditi}%
  \BibitemOpen
  \bibfield  {author} {\bibinfo {author} {\bibfnamefont {A.}~\bibnamefont
  {Sen(De)}}\ and\ \bibinfo {author} {\bibfnamefont {U.}~\bibnamefont {Sen}},\
  }\href {https://doi.org/10.1103/PhysRevA.81.012308} {\bibfield  {journal}
  {\bibinfo  {journal} {Phys. Rev. A}\ }\textbf {\bibinfo {volume} {81}},\
  \bibinfo {pages} {012308} (\bibinfo {year} {2010})}\BibitemShut {NoStop}%
\bibitem [{\citenamefont {Vidal}\ and\ \citenamefont {Werner}(2002)}]{neg4}%
  \BibitemOpen
  \bibfield  {author} {\bibinfo {author} {\bibfnamefont {G.}~\bibnamefont
  {Vidal}}\ and\ \bibinfo {author} {\bibfnamefont {R.~F.}\ \bibnamefont
  {Werner}},\ }\href {https://doi.org/10.1103/PhysRevA.65.032314} {\bibfield
  {journal} {\bibinfo  {journal} {Phys. Rev. A}\ }\textbf {\bibinfo {volume}
  {65}},\ \bibinfo {pages} {032314} (\bibinfo {year} {2002})}\BibitemShut
  {NoStop}%
\bibitem [{\citenamefont {Plenio}(2005)}]{neg5}%
  \BibitemOpen
  \bibfield  {author} {\bibinfo {author} {\bibfnamefont {M.~B.}\ \bibnamefont
  {Plenio}},\ }\href {https://doi.org/10.1103/PhysRevLett.95.090503} {\bibfield
   {journal} {\bibinfo  {journal} {Phys. Rev. Lett.}\ }\textbf {\bibinfo
  {volume} {95}},\ \bibinfo {pages} {090503} (\bibinfo {year}
  {2005})}\BibitemShut {NoStop}%
\bibitem [{\citenamefont {Zanardi}\ \emph {et~al.}(2007)\citenamefont
  {Zanardi}, \citenamefont {Campos~Venuti},\ and\ \citenamefont
  {Giorda}}]{Zanardi2007}%
  \BibitemOpen
  \bibfield  {author} {\bibinfo {author} {\bibfnamefont {P.}~\bibnamefont
  {Zanardi}}, \bibinfo {author} {\bibfnamefont {L.}~\bibnamefont
  {Campos~Venuti}},\ and\ \bibinfo {author} {\bibfnamefont {P.}~\bibnamefont
  {Giorda}},\ }\href {https://doi.org/10.1103/PhysRevA.76.062318} {\bibfield
  {journal} {\bibinfo  {journal} {Phys. Rev. A}\ }\textbf {\bibinfo {volume}
  {76}},\ \bibinfo {pages} {062318} (\bibinfo {year} {2007})}\BibitemShut
  {NoStop}%
\bibitem [{\citenamefont {Satzinger}\ \emph {et~al.}(2021)\citenamefont
  {Satzinger}, \citenamefont {Liu}, \citenamefont {Smith}, \citenamefont
  {Knapp}, \citenamefont {Newman}, \citenamefont {Jones}, \citenamefont {Chen},
  \citenamefont {Quintana}, \citenamefont {Mi}, \citenamefont {Dunsworth},
  \citenamefont {Gidney}, \citenamefont {Aleiner}, \citenamefont {Arute},
  \citenamefont {Arya}, \citenamefont {Atalaya}, \citenamefont {Babbush},
  \citenamefont {Bardin}, \citenamefont {Barends}, \citenamefont {Basso},
  \citenamefont {Bengtsson}, \citenamefont {Bilmes}, \citenamefont {Broughton},
  \citenamefont {Buckley}, \citenamefont {Buell}, \citenamefont {Burkett},
  \citenamefont {Bushnell}, \citenamefont {Chiaro}, \citenamefont {Collins},
  \citenamefont {Courtney}, \citenamefont {Demura}, \citenamefont {Derk},
  \citenamefont {Eppens}, \citenamefont {Erickson}, \citenamefont {Faoro},
  \citenamefont {Farhi}, \citenamefont {Fowler}, \citenamefont {Foxen},
  \citenamefont {Giustina}, \citenamefont {Greene}, \citenamefont {Gross},
  \citenamefont {Harrigan}, \citenamefont {Harrington}, \citenamefont {Hilton},
  \citenamefont {Hong}, \citenamefont {Huang}, \citenamefont {Huggins},
  \citenamefont {Ioffe}, \citenamefont {Isakov}, \citenamefont {Jeffrey},
  \citenamefont {Jiang}, \citenamefont {Kafri}, \citenamefont {Kechedzhi},
  \citenamefont {Khattar}, \citenamefont {Kim}, \citenamefont {Klimov},
  \citenamefont {Korotkov}, \citenamefont {Kostritsa}, \citenamefont
  {Landhuis}, \citenamefont {Laptev}, \citenamefont {Locharla}, \citenamefont
  {Lucero}, \citenamefont {Martin}, \citenamefont {McClean}, \citenamefont
  {McEwen}, \citenamefont {Miao}, \citenamefont {Mohseni}, \citenamefont
  {Montazeri}, \citenamefont {Mruczkiewicz}, \citenamefont {Mutus},
  \citenamefont {Naaman}, \citenamefont {Neeley}, \citenamefont {Neill},
  \citenamefont {Niu}, \citenamefont {O’Brien}, \citenamefont {Opremcak},
  \citenamefont {Pató}, \citenamefont {Petukhov}, \citenamefont {Rubin},
  \citenamefont {Sank}, \citenamefont {Shvarts}, \citenamefont {Strain},
  \citenamefont {Szalay}, \citenamefont {Villalonga}, \citenamefont {White},
  \citenamefont {Yao}, \citenamefont {Yeh}, \citenamefont {Yoo}, \citenamefont
  {Zalcman}, \citenamefont {Neven}, \citenamefont {Boixo}, \citenamefont
  {Megrant}, \citenamefont {Chen}, \citenamefont {Kelly}, \citenamefont
  {Smelyanskiy}, \citenamefont {Kitaev}, \citenamefont {Knap}, \citenamefont
  {Pollmann},\ and\ \citenamefont {Roushan}}]{Satzinger}%
  \BibitemOpen
  \bibfield  {author} {\bibinfo {author} {\bibfnamefont {K.~J.}\ \bibnamefont
  {Satzinger}}, \bibinfo {author} {\bibfnamefont {Y.-J.}\ \bibnamefont {Liu}},
  \bibinfo {author} {\bibfnamefont {A.}~\bibnamefont {Smith}}, \bibinfo
  {author} {\bibfnamefont {C.}~\bibnamefont {Knapp}}, \bibinfo {author}
  {\bibfnamefont {M.}~\bibnamefont {Newman}}, \bibinfo {author} {\bibfnamefont
  {C.}~\bibnamefont {Jones}}, \bibinfo {author} {\bibfnamefont
  {Z.}~\bibnamefont {Chen}}, \bibinfo {author} {\bibfnamefont {C.}~\bibnamefont
  {Quintana}}, \bibinfo {author} {\bibfnamefont {X.}~\bibnamefont {Mi}},
  \bibinfo {author} {\bibfnamefont {A.}~\bibnamefont {Dunsworth}}, \bibinfo
  {author} {\bibfnamefont {C.}~\bibnamefont {Gidney}}, \bibinfo {author}
  {\bibfnamefont {I.}~\bibnamefont {Aleiner}}, \bibinfo {author} {\bibfnamefont
  {F.}~\bibnamefont {Arute}}, \bibinfo {author} {\bibfnamefont
  {K.}~\bibnamefont {Arya}}, \bibinfo {author} {\bibfnamefont {J.}~\bibnamefont
  {Atalaya}}, \bibinfo {author} {\bibfnamefont {R.}~\bibnamefont {Babbush}},
  \bibinfo {author} {\bibfnamefont {J.~C.}\ \bibnamefont {Bardin}}, \bibinfo
  {author} {\bibfnamefont {R.}~\bibnamefont {Barends}}, \bibinfo {author}
  {\bibfnamefont {J.}~\bibnamefont {Basso}}, \bibinfo {author} {\bibfnamefont
  {A.}~\bibnamefont {Bengtsson}}, \bibinfo {author} {\bibfnamefont
  {A.}~\bibnamefont {Bilmes}}, \bibinfo {author} {\bibfnamefont
  {M.}~\bibnamefont {Broughton}}, \bibinfo {author} {\bibfnamefont {B.~B.}\
  \bibnamefont {Buckley}}, \bibinfo {author} {\bibfnamefont {D.~A.}\
  \bibnamefont {Buell}}, \bibinfo {author} {\bibfnamefont {B.}~\bibnamefont
  {Burkett}}, \bibinfo {author} {\bibfnamefont {N.}~\bibnamefont {Bushnell}},
  \bibinfo {author} {\bibfnamefont {B.}~\bibnamefont {Chiaro}}, \bibinfo
  {author} {\bibfnamefont {R.}~\bibnamefont {Collins}}, \bibinfo {author}
  {\bibfnamefont {W.}~\bibnamefont {Courtney}}, \bibinfo {author}
  {\bibfnamefont {S.}~\bibnamefont {Demura}}, \bibinfo {author} {\bibfnamefont
  {A.~R.}\ \bibnamefont {Derk}}, \bibinfo {author} {\bibfnamefont
  {D.}~\bibnamefont {Eppens}}, \bibinfo {author} {\bibfnamefont
  {C.}~\bibnamefont {Erickson}}, \bibinfo {author} {\bibfnamefont
  {L.}~\bibnamefont {Faoro}}, \bibinfo {author} {\bibfnamefont
  {E.}~\bibnamefont {Farhi}}, \bibinfo {author} {\bibfnamefont {A.~G.}\
  \bibnamefont {Fowler}}, \bibinfo {author} {\bibfnamefont {B.}~\bibnamefont
  {Foxen}}, \bibinfo {author} {\bibfnamefont {M.}~\bibnamefont {Giustina}},
  \bibinfo {author} {\bibfnamefont {A.}~\bibnamefont {Greene}}, \bibinfo
  {author} {\bibfnamefont {J.~A.}\ \bibnamefont {Gross}}, \bibinfo {author}
  {\bibfnamefont {M.~P.}\ \bibnamefont {Harrigan}}, \bibinfo {author}
  {\bibfnamefont {S.~D.}\ \bibnamefont {Harrington}}, \bibinfo {author}
  {\bibfnamefont {J.}~\bibnamefont {Hilton}}, \bibinfo {author} {\bibfnamefont
  {S.}~\bibnamefont {Hong}}, \bibinfo {author} {\bibfnamefont {T.}~\bibnamefont
  {Huang}}, \bibinfo {author} {\bibfnamefont {W.~J.}\ \bibnamefont {Huggins}},
  \bibinfo {author} {\bibfnamefont {L.~B.}\ \bibnamefont {Ioffe}}, \bibinfo
  {author} {\bibfnamefont {S.~V.}\ \bibnamefont {Isakov}}, \bibinfo {author}
  {\bibfnamefont {E.}~\bibnamefont {Jeffrey}}, \bibinfo {author} {\bibfnamefont
  {Z.}~\bibnamefont {Jiang}}, \bibinfo {author} {\bibfnamefont
  {D.}~\bibnamefont {Kafri}}, \bibinfo {author} {\bibfnamefont
  {K.}~\bibnamefont {Kechedzhi}}, \bibinfo {author} {\bibfnamefont
  {T.}~\bibnamefont {Khattar}}, \bibinfo {author} {\bibfnamefont
  {S.}~\bibnamefont {Kim}}, \bibinfo {author} {\bibfnamefont {P.~V.}\
  \bibnamefont {Klimov}}, \bibinfo {author} {\bibfnamefont {A.~N.}\
  \bibnamefont {Korotkov}}, \bibinfo {author} {\bibfnamefont {F.}~\bibnamefont
  {Kostritsa}}, \bibinfo {author} {\bibfnamefont {D.}~\bibnamefont {Landhuis}},
  \bibinfo {author} {\bibfnamefont {P.}~\bibnamefont {Laptev}}, \bibinfo
  {author} {\bibfnamefont {A.}~\bibnamefont {Locharla}}, \bibinfo {author}
  {\bibfnamefont {E.}~\bibnamefont {Lucero}}, \bibinfo {author} {\bibfnamefont
  {O.}~\bibnamefont {Martin}}, \bibinfo {author} {\bibfnamefont {J.~R.}\
  \bibnamefont {McClean}}, \bibinfo {author} {\bibfnamefont {M.}~\bibnamefont
  {McEwen}}, \bibinfo {author} {\bibfnamefont {K.~C.}\ \bibnamefont {Miao}},
  \bibinfo {author} {\bibfnamefont {M.}~\bibnamefont {Mohseni}}, \bibinfo
  {author} {\bibfnamefont {S.}~\bibnamefont {Montazeri}}, \bibinfo {author}
  {\bibfnamefont {W.}~\bibnamefont {Mruczkiewicz}}, \bibinfo {author}
  {\bibfnamefont {J.}~\bibnamefont {Mutus}}, \bibinfo {author} {\bibfnamefont
  {O.}~\bibnamefont {Naaman}}, \bibinfo {author} {\bibfnamefont
  {M.}~\bibnamefont {Neeley}}, \bibinfo {author} {\bibfnamefont
  {C.}~\bibnamefont {Neill}}, \bibinfo {author} {\bibfnamefont {M.~Y.}\
  \bibnamefont {Niu}}, \bibinfo {author} {\bibfnamefont {T.~E.}\ \bibnamefont
  {O’Brien}}, \bibinfo {author} {\bibfnamefont {A.}~\bibnamefont {Opremcak}},
  \bibinfo {author} {\bibfnamefont {B.}~\bibnamefont {Pató}}, \bibinfo
  {author} {\bibfnamefont {A.}~\bibnamefont {Petukhov}}, \bibinfo {author}
  {\bibfnamefont {N.~C.}\ \bibnamefont {Rubin}}, \bibinfo {author}
  {\bibfnamefont {D.}~\bibnamefont {Sank}}, \bibinfo {author} {\bibfnamefont
  {V.}~\bibnamefont {Shvarts}}, \bibinfo {author} {\bibfnamefont
  {D.}~\bibnamefont {Strain}}, \bibinfo {author} {\bibfnamefont
  {M.}~\bibnamefont {Szalay}}, \bibinfo {author} {\bibfnamefont
  {B.}~\bibnamefont {Villalonga}}, \bibinfo {author} {\bibfnamefont {T.~C.}\
  \bibnamefont {White}}, \bibinfo {author} {\bibfnamefont {Z.}~\bibnamefont
  {Yao}}, \bibinfo {author} {\bibfnamefont {P.}~\bibnamefont {Yeh}}, \bibinfo
  {author} {\bibfnamefont {J.}~\bibnamefont {Yoo}}, \bibinfo {author}
  {\bibfnamefont {A.}~\bibnamefont {Zalcman}}, \bibinfo {author} {\bibfnamefont
  {H.}~\bibnamefont {Neven}}, \bibinfo {author} {\bibfnamefont
  {S.}~\bibnamefont {Boixo}}, \bibinfo {author} {\bibfnamefont
  {A.}~\bibnamefont {Megrant}}, \bibinfo {author} {\bibfnamefont
  {Y.}~\bibnamefont {Chen}}, \bibinfo {author} {\bibfnamefont {J.}~\bibnamefont
  {Kelly}}, \bibinfo {author} {\bibfnamefont {V.}~\bibnamefont {Smelyanskiy}},
  \bibinfo {author} {\bibfnamefont {A.}~\bibnamefont {Kitaev}}, \bibinfo
  {author} {\bibfnamefont {M.}~\bibnamefont {Knap}}, \bibinfo {author}
  {\bibfnamefont {F.}~\bibnamefont {Pollmann}},\ and\ \bibinfo {author}
  {\bibfnamefont {P.}~\bibnamefont {Roushan}},\ }\href
  {https://doi.org/10.1126/science.abi8378} {\bibfield  {journal} {\bibinfo
  {journal} {Science}\ }\textbf {\bibinfo {volume} {374}},\ \bibinfo {pages}
  {1237} (\bibinfo {year} {2021})},\ \Eprint
  {https://arxiv.org/abs/https://www.science.org/doi/pdf/10.1126/science.abi8378}
  {https://www.science.org/doi/pdf/10.1126/science.abi8378} \BibitemShut
  {NoStop}%
\bibitem [{\citenamefont {Verresen}\ \emph {et~al.}(2021)\citenamefont
  {Verresen}, \citenamefont {Lukin},\ and\ \citenamefont
  {Vishwanath}}]{Vishwanath_PRX}%
  \BibitemOpen
  \bibfield  {author} {\bibinfo {author} {\bibfnamefont {R.}~\bibnamefont
  {Verresen}}, \bibinfo {author} {\bibfnamefont {M.~D.}\ \bibnamefont
  {Lukin}},\ and\ \bibinfo {author} {\bibfnamefont {A.}~\bibnamefont
  {Vishwanath}},\ }\href {https://doi.org/10.1103/PhysRevX.11.031005}
  {\bibfield  {journal} {\bibinfo  {journal} {Phys. Rev. X}\ }\textbf {\bibinfo
  {volume} {11}},\ \bibinfo {pages} {031005} (\bibinfo {year}
  {2021})}\BibitemShut {NoStop}%
\bibitem [{\citenamefont {Semeghini}\ \emph {et~al.}(2021)\citenamefont
  {Semeghini}, \citenamefont {Levine}, \citenamefont {Keesling}, \citenamefont
  {Ebadi}, \citenamefont {Wang}, \citenamefont {Bluvstein}, \citenamefont
  {Verresen}, \citenamefont {Pichler}, \citenamefont {Kalinowski},
  \citenamefont {Samajdar}, \citenamefont {Omran}, \citenamefont {Sachdev},
  \citenamefont {Vishwanath}, \citenamefont {Greiner}, \citenamefont
  {Vuletić},\ and\ \citenamefont {Lukin}}]{Vishwanath_science}%
  \BibitemOpen
  \bibfield  {author} {\bibinfo {author} {\bibfnamefont {G.}~\bibnamefont
  {Semeghini}}, \bibinfo {author} {\bibfnamefont {H.}~\bibnamefont {Levine}},
  \bibinfo {author} {\bibfnamefont {A.}~\bibnamefont {Keesling}}, \bibinfo
  {author} {\bibfnamefont {S.}~\bibnamefont {Ebadi}}, \bibinfo {author}
  {\bibfnamefont {T.~T.}\ \bibnamefont {Wang}}, \bibinfo {author}
  {\bibfnamefont {D.}~\bibnamefont {Bluvstein}}, \bibinfo {author}
  {\bibfnamefont {R.}~\bibnamefont {Verresen}}, \bibinfo {author}
  {\bibfnamefont {H.}~\bibnamefont {Pichler}}, \bibinfo {author} {\bibfnamefont
  {M.}~\bibnamefont {Kalinowski}}, \bibinfo {author} {\bibfnamefont
  {R.}~\bibnamefont {Samajdar}}, \bibinfo {author} {\bibfnamefont
  {A.}~\bibnamefont {Omran}}, \bibinfo {author} {\bibfnamefont
  {S.}~\bibnamefont {Sachdev}}, \bibinfo {author} {\bibfnamefont
  {A.}~\bibnamefont {Vishwanath}}, \bibinfo {author} {\bibfnamefont
  {M.}~\bibnamefont {Greiner}}, \bibinfo {author} {\bibfnamefont
  {V.}~\bibnamefont {Vuletić}},\ and\ \bibinfo {author} {\bibfnamefont
  {M.~D.}\ \bibnamefont {Lukin}},\ }\href
  {https://doi.org/10.1126/science.abi8794} {\bibfield  {journal} {\bibinfo
  {journal} {Science}\ }\textbf {\bibinfo {volume} {374}},\ \bibinfo {pages}
  {1242} (\bibinfo {year} {2021})},\ \Eprint
  {https://arxiv.org/abs/https://www.science.org/doi/pdf/10.1126/science.abi8794}
  {https://www.science.org/doi/pdf/10.1126/science.abi8794} \BibitemShut
  {NoStop}%
\bibitem [{\citenamefont {Auger}\ \emph {et~al.}(2017)\citenamefont {Auger},
  \citenamefont {Bergamini},\ and\ \citenamefont {Browne}}]{Auger}%
  \BibitemOpen
  \bibfield  {author} {\bibinfo {author} {\bibfnamefont {J.~M.}\ \bibnamefont
  {Auger}}, \bibinfo {author} {\bibfnamefont {S.}~\bibnamefont {Bergamini}},\
  and\ \bibinfo {author} {\bibfnamefont {D.~E.}\ \bibnamefont {Browne}},\
  }\href {https://doi.org/10.1103/PhysRevA.96.052320} {\bibfield  {journal}
  {\bibinfo  {journal} {Phys. Rev. A}\ }\textbf {\bibinfo {volume} {96}},\
  \bibinfo {pages} {052320} (\bibinfo {year} {2017})}\BibitemShut {NoStop}%
\bibitem [{\citenamefont {Pachos}\ \emph {et~al.}(2009)\citenamefont {Pachos},
  \citenamefont {Wieczorek}, \citenamefont {Schmid}, \citenamefont {Kiesel},
  \citenamefont {Pohlner},\ and\ \citenamefont {Weinfurter}}]{Pachos_2009}%
  \BibitemOpen
  \bibfield  {author} {\bibinfo {author} {\bibfnamefont {J.~K.}\ \bibnamefont
  {Pachos}}, \bibinfo {author} {\bibfnamefont {W.}~\bibnamefont {Wieczorek}},
  \bibinfo {author} {\bibfnamefont {C.}~\bibnamefont {Schmid}}, \bibinfo
  {author} {\bibfnamefont {N.}~\bibnamefont {Kiesel}}, \bibinfo {author}
  {\bibfnamefont {R.}~\bibnamefont {Pohlner}},\ and\ \bibinfo {author}
  {\bibfnamefont {H.}~\bibnamefont {Weinfurter}},\ }\href
  {https://doi.org/10.1088/1367-2630/11/8/083010} {\bibfield  {journal}
  {\bibinfo  {journal} {New Journal of Physics}\ }\textbf {\bibinfo {volume}
  {11}},\ \bibinfo {pages} {083010} (\bibinfo {year} {2009})}\BibitemShut
  {NoStop}%
\bibitem [{\citenamefont {Tsomokos}\ \emph {et~al.}(2009)\citenamefont
  {Tsomokos}, \citenamefont {Hamma}, \citenamefont {Zhang}, \citenamefont
  {Haas},\ and\ \citenamefont {Fazio}}]{Fazio2009}%
  \BibitemOpen
  \bibfield  {author} {\bibinfo {author} {\bibfnamefont {D.~I.}\ \bibnamefont
  {Tsomokos}}, \bibinfo {author} {\bibfnamefont {A.}~\bibnamefont {Hamma}},
  \bibinfo {author} {\bibfnamefont {W.}~\bibnamefont {Zhang}}, \bibinfo
  {author} {\bibfnamefont {S.}~\bibnamefont {Haas}},\ and\ \bibinfo {author}
  {\bibfnamefont {R.}~\bibnamefont {Fazio}},\ }\href
  {https://doi.org/10.1103/PhysRevA.80.060302} {\bibfield  {journal} {\bibinfo
  {journal} {Phys. Rev. A}\ }\textbf {\bibinfo {volume} {80}},\ \bibinfo
  {pages} {060302} (\bibinfo {year} {2009})}\BibitemShut {NoStop}%
\bibitem [{\citenamefont {Vajna}\ and\ \citenamefont {D\'ora}(2014)}]{Vajna14}%
  \BibitemOpen
  \bibfield  {author} {\bibinfo {author} {\bibfnamefont {S.}~\bibnamefont
  {Vajna}}\ and\ \bibinfo {author} {\bibfnamefont {B.}~\bibnamefont {D\'ora}},\
  }\href {https://doi.org/10.1103/PhysRevB.89.161105} {\bibfield  {journal}
  {\bibinfo  {journal} {Phys. Rev. B}\ }\textbf {\bibinfo {volume} {89}},\
  \bibinfo {pages} {161105} (\bibinfo {year} {2014})}\BibitemShut {NoStop}%
\bibitem [{\citenamefont {Andraschko}\ and\ \citenamefont
  {Sirker}(2014)}]{Sirker14}%
  \BibitemOpen
  \bibfield  {author} {\bibinfo {author} {\bibfnamefont {F.}~\bibnamefont
  {Andraschko}}\ and\ \bibinfo {author} {\bibfnamefont {J.}~\bibnamefont
  {Sirker}},\ }\href {https://doi.org/10.1103/PhysRevB.89.125120} {\bibfield
  {journal} {\bibinfo  {journal} {Phys. Rev. B}\ }\textbf {\bibinfo {volume}
  {89}},\ \bibinfo {pages} {125120} (\bibinfo {year} {2014})}\BibitemShut
  {NoStop}%
\bibitem [{\citenamefont {Gurarie}(2019)}]{gurarie_pra_2019}%
  \BibitemOpen
  \bibfield  {author} {\bibinfo {author} {\bibfnamefont {V.}~\bibnamefont
  {Gurarie}},\ }\href {https://doi.org/10.1103/PhysRevA.100.031601} {\bibfield
  {journal} {\bibinfo  {journal} {Phys. Rev. A}\ }\textbf {\bibinfo {volume}
  {100}},\ \bibinfo {pages} {031601} (\bibinfo {year} {2019})}\BibitemShut
  {NoStop}%
\bibitem [{\citenamefont {Nandi}\ \emph {et~al.}(2022)\citenamefont {Nandi},
  \citenamefont {Bhattacharyya},\ and\ \citenamefont
  {Dasgupta}}]{nandi_prl_2022}%
  \BibitemOpen
  \bibfield  {author} {\bibinfo {author} {\bibfnamefont {P.}~\bibnamefont
  {Nandi}}, \bibinfo {author} {\bibfnamefont {S.}~\bibnamefont
  {Bhattacharyya}},\ and\ \bibinfo {author} {\bibfnamefont {S.}~\bibnamefont
  {Dasgupta}},\ }\href {https://doi.org/10.1103/PhysRevLett.128.247201}
  {\bibfield  {journal} {\bibinfo  {journal} {Phys. Rev. Lett.}\ }\textbf
  {\bibinfo {volume} {128}},\ \bibinfo {pages} {247201} (\bibinfo {year}
  {2022})}\BibitemShut {NoStop}%
\bibitem [{\citenamefont {Jafari}(2019)}]{Jafari_dqpt1}%
  \BibitemOpen
  \bibfield  {author} {\bibinfo {author} {\bibfnamefont {R.}~\bibnamefont
  {Jafari}},\ }\href {https://doi.org/10.1038/s41598-019-39595-3} {\bibfield
  {journal} {\bibinfo  {journal} {Scientific Reports}\ }\textbf {\bibinfo
  {volume} {9}},\ \bibinfo {pages} {2871} (\bibinfo {year} {2019})}\BibitemShut
  {NoStop}%
\bibitem [{\citenamefont {Wei}\ and\ \citenamefont {Goldbart}(2003)}]{ggm1}%
  \BibitemOpen
  \bibfield  {author} {\bibinfo {author} {\bibfnamefont {T.-C.}\ \bibnamefont
  {Wei}}\ and\ \bibinfo {author} {\bibfnamefont {P.~M.}\ \bibnamefont
  {Goldbart}},\ }\href {https://doi.org/10.1103/PhysRevA.68.042307} {\bibfield
  {journal} {\bibinfo  {journal} {Phys. Rev. A}\ }\textbf {\bibinfo {volume}
  {68}},\ \bibinfo {pages} {042307} (\bibinfo {year} {2003})}\BibitemShut
  {NoStop}%
\bibitem [{\citenamefont {Blasone}\ \emph {et~al.}(2008)\citenamefont
  {Blasone}, \citenamefont {Dell'Anno}, \citenamefont {De~Siena},\ and\
  \citenamefont {Illuminati}}]{ggm3}%
  \BibitemOpen
  \bibfield  {author} {\bibinfo {author} {\bibfnamefont {M.}~\bibnamefont
  {Blasone}}, \bibinfo {author} {\bibfnamefont {F.}~\bibnamefont {Dell'Anno}},
  \bibinfo {author} {\bibfnamefont {S.}~\bibnamefont {De~Siena}},\ and\
  \bibinfo {author} {\bibfnamefont {F.}~\bibnamefont {Illuminati}},\ }\href
  {https://doi.org/10.1103/PhysRevA.77.062304} {\bibfield  {journal} {\bibinfo
  {journal} {Phys. Rev. A}\ }\textbf {\bibinfo {volume} {77}},\ \bibinfo
  {pages} {062304} (\bibinfo {year} {2008})}\BibitemShut {NoStop}%
\bibitem [{\citenamefont {Or\'us}(2008)}]{ggm4}%
  \BibitemOpen
  \bibfield  {author} {\bibinfo {author} {\bibfnamefont {R.}~\bibnamefont
  {Or\'us}},\ }\href {https://doi.org/10.1103/PhysRevA.78.062332} {\bibfield
  {journal} {\bibinfo  {journal} {Phys. Rev. A}\ }\textbf {\bibinfo {volume}
  {78}},\ \bibinfo {pages} {062332} (\bibinfo {year} {2008})}\BibitemShut
  {NoStop}%
\bibitem [{\citenamefont {Buchholz}\ \emph {et~al.}(2016)\citenamefont
  {Buchholz}, \citenamefont {Moroder},\ and\ \citenamefont
  {Gühne}}]{ggm_mixed_otfried}%
  \BibitemOpen
  \bibfield  {author} {\bibinfo {author} {\bibfnamefont {L.~E.}\ \bibnamefont
  {Buchholz}}, \bibinfo {author} {\bibfnamefont {T.}~\bibnamefont {Moroder}},\
  and\ \bibinfo {author} {\bibfnamefont {O.}~\bibnamefont {Gühne}},\ }\href
  {https://doi.org/https://doi.org/10.1002/andp.201500293} {\bibfield
  {journal} {\bibinfo  {journal} {Annalen der Physik}\ }\textbf {\bibinfo
  {volume} {528}},\ \bibinfo {pages} {278} (\bibinfo {year} {2016})},\ \Eprint
  {https://arxiv.org/abs/https://arxiv.org/pdf/1412.7471.pdf}
  {https://arxiv.org/pdf/1412.7471.pdf} \BibitemShut {NoStop}%
\bibitem [{\citenamefont {Das}\ \emph {et~al.}(2016)\citenamefont {Das},
  \citenamefont {Roy}, \citenamefont {Bagchi}, \citenamefont {Misra},
  \citenamefont {Sen(De)},\ and\ \citenamefont {Sen}}]{aditi_ggm_mixed}%
  \BibitemOpen
  \bibfield  {author} {\bibinfo {author} {\bibfnamefont {T.}~\bibnamefont
  {Das}}, \bibinfo {author} {\bibfnamefont {S.~S.}\ \bibnamefont {Roy}},
  \bibinfo {author} {\bibfnamefont {S.}~\bibnamefont {Bagchi}}, \bibinfo
  {author} {\bibfnamefont {A.}~\bibnamefont {Misra}}, \bibinfo {author}
  {\bibfnamefont {A.}~\bibnamefont {Sen(De)}},\ and\ \bibinfo {author}
  {\bibfnamefont {U.}~\bibnamefont {Sen}},\ }\href
  {https://doi.org/10.1103/PhysRevA.94.022336} {\bibfield  {journal} {\bibinfo
  {journal} {Phys. Rev. A}\ }\textbf {\bibinfo {volume} {94}},\ \bibinfo
  {pages} {022336} (\bibinfo {year} {2016})}\BibitemShut {NoStop}%
\bibitem [{\citenamefont {Peres}(1996)}]{peres}%
  \BibitemOpen
  \bibfield  {author} {\bibinfo {author} {\bibfnamefont {A.}~\bibnamefont
  {Peres}},\ }\href {https://doi.org/10.1103/PhysRevLett.77.1413} {\bibfield
  {journal} {\bibinfo  {journal} {Phys. Rev. Lett.}\ }\textbf {\bibinfo
  {volume} {77}},\ \bibinfo {pages} {1413} (\bibinfo {year}
  {1996})}\BibitemShut {NoStop}%
\bibitem [{\citenamefont {Horodecki}\ \emph {et~al.}(1996)\citenamefont
  {Horodecki}, \citenamefont {Horodecki},\ and\ \citenamefont
  {Horodecki}}]{horodecki1996}%
  \BibitemOpen
  \bibfield  {author} {\bibinfo {author} {\bibfnamefont {M.}~\bibnamefont
  {Horodecki}}, \bibinfo {author} {\bibfnamefont {P.}~\bibnamefont
  {Horodecki}},\ and\ \bibinfo {author} {\bibfnamefont {R.}~\bibnamefont
  {Horodecki}},\ }\href
  {https://doi.org/https://doi.org/10.1016/S0375-9601(96)00706-2} {\bibfield
  {journal} {\bibinfo  {journal} {Physics Letters A}\ }\textbf {\bibinfo
  {volume} {223}},\ \bibinfo {pages} {1} (\bibinfo {year} {1996})}\BibitemShut
  {NoStop}%
\bibitem [{\citenamefont {Dhahri}(2008)}]{dhahri2008lindblad}%
  \BibitemOpen
  \bibfield  {author} {\bibinfo {author} {\bibfnamefont {A.}~\bibnamefont
  {Dhahri}},\ }\href@noop {} {\bibfield  {journal} {\bibinfo  {journal}
  {Journal of Physics A: Mathematical and Theoretical}\ }\textbf {\bibinfo
  {volume} {41}},\ \bibinfo {pages} {275305} (\bibinfo {year}
  {2008})}\BibitemShut {NoStop}%
\bibitem [{\citenamefont {Chanda}\ \emph {et~al.}(2018)\citenamefont {Chanda},
  \citenamefont {Das}, \citenamefont {Sadhukhan}, \citenamefont {Pal},
  \citenamefont {Sen(De)},\ and\ \citenamefont {Sen}}]{Titas18}%
  \BibitemOpen
  \bibfield  {author} {\bibinfo {author} {\bibfnamefont {T.}~\bibnamefont
  {Chanda}}, \bibinfo {author} {\bibfnamefont {T.}~\bibnamefont {Das}},
  \bibinfo {author} {\bibfnamefont {D.}~\bibnamefont {Sadhukhan}}, \bibinfo
  {author} {\bibfnamefont {A.~K.}\ \bibnamefont {Pal}}, \bibinfo {author}
  {\bibfnamefont {A.}~\bibnamefont {Sen(De)}},\ and\ \bibinfo {author}
  {\bibfnamefont {U.}~\bibnamefont {Sen}},\ }\href
  {https://doi.org/10.1103/PhysRevA.97.062324} {\bibfield  {journal} {\bibinfo
  {journal} {Phys. Rev. A}\ }\textbf {\bibinfo {volume} {97}},\ \bibinfo
  {pages} {062324} (\bibinfo {year} {2018})}\BibitemShut {NoStop}%
\bibitem [{\citenamefont {Ángel Rivas}\ \emph {et~al.}(2014)\citenamefont
  {Ángel Rivas}, \citenamefont {Huelga},\ and\ \citenamefont
  {Plenio}}]{Rivas_2014}%
  \BibitemOpen
  \bibfield  {author} {\bibinfo {author} {\bibnamefont {Ángel Rivas}},
  \bibinfo {author} {\bibfnamefont {S.~F.}\ \bibnamefont {Huelga}},\ and\
  \bibinfo {author} {\bibfnamefont {M.~B.}\ \bibnamefont {Plenio}},\ }\href
  {https://doi.org/10.1088/0034-4885/77/9/094001} {\bibfield  {journal}
  {\bibinfo  {journal} {Reports on Progress in Physics}\ }\textbf {\bibinfo
  {volume} {77}},\ \bibinfo {pages} {094001} (\bibinfo {year}
  {2014})}\BibitemShut {NoStop}%
\bibitem [{\citenamefont {Breuer}\ and\ \citenamefont
  {Petruccione}(2007)}]{noise1}%
  \BibitemOpen
  \bibfield  {author} {\bibinfo {author} {\bibfnamefont {H.-P.}\ \bibnamefont
  {Breuer}}\ and\ \bibinfo {author} {\bibfnamefont {F.}~\bibnamefont
  {Petruccione}},\ }\href
  {https://doi.org/10.1093/acprof:oso/9780199213900.001.0001} {\emph {\bibinfo
  {title} {The Theory of Open Quantum Systems}}}\ (\bibinfo  {publisher}
  {Oxford University Press},\ \bibinfo {year} {2007})\BibitemShut {NoStop}%
\bibitem [{\citenamefont {Rivas}\ and\ \citenamefont
  {Huelga}(2012)}]{noisebook}%
  \BibitemOpen
  \bibfield  {author} {\bibinfo {author} {\bibfnamefont {A.}~\bibnamefont
  {Rivas}}\ and\ \bibinfo {author} {\bibfnamefont {S.~F.}\ \bibnamefont
  {Huelga}},\ }\href {https://doi.org/10.1007/978-3-642-23354-8} {\emph
  {\bibinfo {title} {Open Quantum Systems}}}\ (\bibinfo  {publisher}
  {Springer-Verlag Berlin Heidelberg},\ \bibinfo {year} {2012})\BibitemShut
  {NoStop}%
\bibitem [{\citenamefont {Reif}(1965)}]{REIF65}%
  \BibitemOpen
  \bibfield  {author} {\bibinfo {author} {\bibfnamefont {F.}~\bibnamefont
  {Reif}},\ }\href@noop {} {\emph {\bibinfo {title} {Fundamentals of
  Statistical and Thermal Physics}}}\ (\bibinfo  {publisher} {McGraw Hill},\
  \bibinfo {address} {Tokyo},\ \bibinfo {year} {1965})\BibitemShut {NoStop}%
\end{thebibliography}%

\end{document}